\documentclass{iopjournal}
\usepackage[utf8]{inputenc}
\usepackage[english]{babel}
\usepackage[T1]{fontenc}
\usepackage{amssymb}
\usepackage{amsmath}
\usepackage{cases}
\usepackage{epsfig}
\usepackage{physics}
\PassOptionsToPackage{colorlinks=true,allcolors=black}{hyperref}
\usepackage{array}
\usepackage[all]{xy}
\usepackage{algorithm}
\usepackage{algpseudocode}
\usepackage{tikz}
\usetikzlibrary{shapes.geometric, arrows}
\hypersetup{
    linkcolor= black,  
    citecolor= black,   
    filecolor= black,   
    urlcolor= black,   
}
\usepackage{multirow}
\usepackage{qcircuit}
\usepackage{lipsum}
\usepackage{color}
\usepackage{qcircuit}
\usepackage[export]{adjustbox}

\begin{document}

\renewcommand{\thefootnote}{\fnsymbol{footnote}}

\articletype{Paper} 

\title{Quantum Dialogue through Non-Destructive Discrimination of Cluster State}

\author{Mandar Thatte$^{1,2}$, Shreya Banerjee$^{3,2}$ and Prasanta. K. Panigrahi$^{2,4}$}

\affil{$^1$ Department of Mathematics, Institute of Chemical Technology, Mumbai-400019, Maharashtra, India.}

\affil{$^2$Center for Quantum Science and Technology, Siksha 'O' Anusandhan, Bhubaneswar-751030, Odisha, India.}

\affil{$^3$ Department of Physics and Astronomy, University of Exeter, Stocker Road, Exeter-EX4 4QL, United Kingdom}

\affil{$^4$Department of Physical Sciences, Indian Institute of Science Education and Research Kolkata, Mohanpur-741246, West Bengal, India}

\email{mat22mm.thatte@pg.ictmumbai.edu.in $^1$, s.banerjee3@exeter.ac.uk $^2$, pprasanta@iiserkol.ac.in $^3$}


\begin{abstract}
 We propose an efficient, measurement-based quantum dialogue protocol through non-destructive discrimination (NDD) of cluster state. We use ancilla-based measurements that allow the state to be reused without destroying its entanglement. The initial state is a local unitary (LU-) equivalent of the five-qubit cluster state, which significantly reduces the state components from $32$ to $4$, simultaneously allowing one to write its different subspaces using two different bases. The protocol utilizes single qubit unitaries from the Pauli group to encode the messages, thus preserving the stabilizer nature of the initial state throughout. We demonstrate that the proposed protocol is secure under common quantum attacks and outlining the procedure for the scalability of the scheme to transmit an $n$-bit message. The proposed protocol has been experimentally verified using IBM quantum backend 'IBM-Torino' as a proof of concept. Using the stabilizer nature of the state, we further introduce a single-qubit error correction mechanism that enhances robustness against noise without requiring any additional qubits. Further, the use of NDD allows one to reuse the quantum resources in advancing the two-way dialogue, marking the importance and novelty of the proposed scheme over preexisting methods.
\end{abstract}

\section{Introduction}
With the current advancements of technologies, secure communication is crucial at every stages of our society, starting from everyday finance sectors to defense of individual countries. At its core, secure communication translates to transmitting information between participants without any leakage or eavesdropping. While cryptographic protocols date back to thousands of years, uses of machines and computers to encrypt messages started in the past century. Recent developments in quantum computing pushes this boundary further, with cryptographic protocols facing new opportunities and challenges \cite{ PhysRevA.111.012603, Shukla2025, zhen-zhen,two-step}. 
In general, in a cryptographic protocol, a shared key is used to encrypt and decrypt certain information. Based on the technologies used, these protocols can be broadly categorized into two groups: Classical and Quantum Cryptography. Quantum cryptography uses principles of quantum mechanics to encrypt classical information and promises unconditional security in an idealistic theoretical scenario. In 1984, Bennet and Brassard proposed the well known BB84 protocol \cite{bennett2014quantum}, initiating the applications of quantum cryptography with Quantum Key Distribution (QKD) protocols \cite{bennett2014quantum,ekert1991quantum, bennett1992quantum, goldenberg1995quantum}. Soon, several experimental realizations of cryptographic protocols \cite{zhou2004quantum, wang2005experimental} followed as technological advancements  were seen in quantum hardware development. In QKD, a key is shared between two or more parties securely with quantum means, to be used in an encryption scheme later. However, it was observed that pre-sharing key is not necessary for secure quantum communication which led to Quantum Secure Direct Communication (QSDC) \cite{zhu2017experimental, long2002theoretically, deng2003two, lee2006quantum, cao2010quantum}. In QSDC, entanglement is used as a resource to transmit information without any pre-shared key.

Quantum Dialogue is another application of quantum cryptography where two parties can exchange information simultaneously using same channel. The first protocol for quantum dialogue was proposed in  \cite{nguyen2004quantum} using Bell state, where, one qubit was home qubit and other was travel qubit. Information would be encoded on the travel qubit for the protocol. However, later, in \cite{zhong2005quantum}, it was pointed out that the protocol proposed in \cite{nguyen2004quantum} is vulnerable towards the intercept-resend attack. The authors also  presented an improved version of it \cite{zhong2005quantum}. Later B. A. Nguyen himself proposed another improved version of quantum dialogue protocol \cite{an2005secure}. Further advances in quantum dialogue led to exploring different approaches. In \cite{li2009quantum, xia2006quantum} the authors proposed quantum dialogue schemes using the W and GHZ states respectively. Dense coding \cite{zhong2005quantum, xia2006quantum, wen2007secure, chauhan2021quantum, CIQD}, Entanglement Swapping \cite{dong2008controlled, gao2010two, ye2013quantum,wang2016efficient}, Single photon
\cite{xin2006secure, shi2010quantum}, Measurement Device Independent \cite{maitra2017measurement, das2020two, zhang2024measurement} etc have also been exploited to develop several quantum dialogue schemes. In \cite{banerjee2017asymmetric}, authors has presented an  asymmetric quantum dialogue protocol, in contrast with the most quantum dialogue schemes present in the literature which are symmetric. Several studies also focused on the vulnerabilities of quantum dialogue schemes. Gao et.al and Tan et.al independently pointed out information leakage or classical correlation in quantum dialogue \cite{gao2008revisiting, tan2008classical}. More work on information leakage have been done in \cite{gao2010two, ye2013quantum, wang2016efficient}. \\ Many quantum dialogue protocols assume that the classical communication channel used in communication is authenticated that is, eavesdropper could listen but not alter the message. In secure communication, it is crucial that the communicating parties authenticate themselves before communication. Without authentication, communication is vulnerable to impersonation attacks. To overcome this challenge, Authenticated quantum dialogue (AQD) protocols \cite{prob, shend, LinCY} have been proposed that integrates authentication mechanism with the dialogue protocols. A pre-shared key is used to verify each other's identities before or during the protocol to prevent impersonation attacks. An authenticated quantum dialogue using Bell states has been proposed in \cite{shend, LinCY}. In \cite{prob}, authors have presented a probabilistic AQD offering probabilistic encoding and key reuse. Semi-quantum dialogue (SQD) protocols \cite{zhen-zhen, Tian, semi} are a special type of protocols where only one party has full quantum capabilities while the other has classical capabilities. The quantum party can prepare or measure quantum superposed states, while the capabilities of classical party are restricted to preparing and measuring in classical basis, reflecting qubits and reordering qubits. SQD protocols based on single photon \cite{Tian}, Bell states \cite{semi}, $\Omega$ state \cite{zhen-zhen} have been proposed. Collective Noises such as collective dephasing noise and collective rotation noise have grabbed significant attention in quantum dialogues. It is observed that some Bell states forms a Decoherence Free Space(DFS) against these collective noises. Using this property, significant quantum dialogues schemes has been proposed to overcome collective noises \cite{two-step, yang2013quantum, ye2014information, ye2022fault, yang2019new, Hzu}.   
A sufficient condition to construct quantum dialogue using a set of quantum states and group of unitary operators was proposed in \cite{shukla2013group}. This provided a group theoretic structure for unitary operations used for dense coding in quantum dialogue, marking an important theoretical invention. The general structure of most of the quantum dialogue scheme between two parties (say Alice and Bob) present in the literature is as follows. Alice prepares an multi-qubit entangled state, and chooses a few qubits of this state to be the travel qubits on which message would be encoded and sent to Bob. The other qubits, namely, the home qubits remains with Alice. Alice encodes her message on the travel qubits and sends them to Bob. After receiving the travel qubits, Bob also encodes his message on those travel qubits and sends them back to Alice. By measuring both home and travel qubits, Alice obtains Bob's encoding and decodes his message. Then, Alice announces the initial state, final state and the measurement outcomes, using these, Bob decodes Alice's message. While doing this, there are chances that Eve might get some information.

Our proposed scheme uses a five-qubit quantum state that is a Local Unitary (LU) equivalent of the five-qubit cluster state. We present a set of $32$ mutually orthogonal states that are also LU equivalent of the five-qubit cluster state. When Alice encodes her message, the initial state transforms into one of these 32 orthogonal states and is then transmitted to Bob. Bob performs non-destructive discrimination (NDD) \cite{Samal_2010}, which without disturbing the state and its entanglement, discriminate itself from its 31 orthogonal counterparts. As for each orthogonal state, NDD provides distinctive output, depending on the NDD outcome, Bob deciphers the transmitted state as well as the unitary operation of Alice, successfully decoding her message. Subsequently, Bob encodes his message in the same transmitted state which again transforms into one of the orthogonal states. Performing NDD on the state Alice receives, she retrieves the encrypted state and since Alice knows the initial state and her encoding operation, Alice deciphers Bob's unitary operation and thus decodes his message. The point to highlight here is, Alice or Bob does not have to announce the states for communication which prevents eavesdropping.

A quantum dialogue scheme was proposed using NDD of Brown state in \cite{jain2009secure}. Non-destructive discrimination promotes the re-usability of the same quantum state for two-way quantum dialogue, increasing its resource efficiency. While our proposed protocol draws inspiration from the scheme presented in \cite{jain2009secure}, we have incorporated the group theoretic structure proposed by \cite{shukla2013group} in the encoding of the message. This marks a key difference between the two dialogue schemes. Additionally, it  specifies a significant improvement from the protocol prescribed in \cite{jain2009secure}. The unitary operations used for encoding in \cite{jain2009secure} fails to follow a group theoretic structure, which can lead to ambiguous encoding outcomes, i.e., different encoding unitaries applied to the initial Brown state can transform it into same orthogonal state.  This results into same NDD outcome for two different encodings. From the perspective of the dialogue problem, this means, two different messages mapped to the same encrypted message. In the proposed protocol, we have replaced Brown state by (LU equivalent) Cluster state and have introduced a group theoretic structure for the encoding unitary operations. This ensures the mapping of two different encoded messages into different orthogonal states. An added advantage of using a (LU equivalent) cluster state, is that it is a stabilizer state, and thus can be useful in detecting and correcting quantum errors. In addition to our quantum dialogue scheme, we have also proposed a single-qubit error correction scheme for our protocol, utilizing the stabilizer nature of cluster states. 
 
The rest of the paper is organized as follows. In Section \ref{sec:2} we briefly discuss about cluster states and non-destructive discrimination (NDD). In Section \ref{sec:3} we provide the encoding scheme used for the protocol. Section \ref{sec:4} explains our proposed protocol for quantum dialogue, our main result. We outline several security aspects of our protocol in Section \ref{sec:5}, followed by its efficiency in Section \ref{sec:6}. In Section \ref{sec:7} we generalize our quantum dialogue scheme for an n-bit message using an n-qubit (LU equivalent) cluster state. In Section \ref{sqec} the single-qubit error correction scheme for our protocol is outlined. We finally conclude in Section \ref{conc} with future directions. 

\section{Preliminary}\label{sec:2}
\subsection{Cluster States}
Cluster states forms a class of multipartite entangled states. The general form of an n-qubit cluster state is: 

$$ \ket {C_{n}} = \frac{1}{2^{n/2}} \otimes_{i=1}^{n} ( \ket{0}_i Z^{i+1} + \ket{1}_i), $$
with, $Z^{n+1}=1.$

\begin{figure}[!hbt]
    \centerline{
    \Qcircuit @C=3.8em @R=.6em{
    & \gate{H} & \ctrl{1} & \qw & \qw & \qw  & \qw \\
    & \gate{H} & \control \qw & \ctrl{1} & \qw & \qw & \qw \\
    & \gate{H} &  \qw & \control \qw & \ctrl{1} & \qw & \qw \\
    & \gate{H} & \qw & \qw & \control \qw & \ctrl{1} & \qw \\
    & \gate{H} & \qw & \qw & \qw & \control \qw & \qw
    }
    }
    \caption{Circuit diagram of five-qubit Cluster state $\ket{C_5}$}
    \label{fig:1}
\end{figure}

This means, starting at $\ket{+}$, there exist a nearest neighbor controlled Z operation, until qubit $n$ becomes the control. For five qubits, this technique yields the circuit provided in Fig. \ref{fig:1}. These states have been experimentally realized in various physical architectures \cite{articleLu, Blythe_2006,articleHaddadi}. As is well known, cluster states are stabilizer states, i.e., there exists operators $S_i$ such that $S_i \ket{C_n} = \ket{C_n}$. This property makes cluster states quite useful in quantum error correction schemes \cite{citation-key}, error protected communication schemes,  as well as encryption protocols \cite{ PhysRevA.78.062333, 9605278}. 

In \cite{Briege}, authors have introduced  local unitary (LU) equivalent states of a cluster state. These states can be obtained by performing only local unitary operations on the cluster state. These states preserve the key properties of the original cluster state, including its entanglement structure and computational capabilities. Using this principle, we can construct a  LU equivalent quantum state $\ket{\psi_5}$ of the five-qubit cluster state, by applying Hadamard gates on the first, third, and fifth qubits. The circuit diagram for LU equivalent $\ket{C_5}$ state is shown in Fig. \ref{fig:2}, and the quantum state is provided in Eq.~\ref{eq:1}. It can be easily seen that the $\ket{\psi_5}$ can also be generated using the circuit provided in Fig. \ref{fig:3}.

\begin{figure}[!hbt]
    \centerline{
    \Qcircuit @C=2.8em @R=.6em{
    & \gate{H} & \ctrl{1} & \qw & \qw   & \qw \barrier{4} & \qw & \gate{H} & \qw \\
    & \gate{H} & \control \qw & \ctrl{1} & \qw & \qw  & \qw & \qw & \qw  \\
    & \gate{H} &  \qw & \control \qw & \ctrl{1} & \qw  & \qw & \gate{H} & \qw  \\
    & \gate{H} & \qw & \qw & \control \qw & \ctrl{1}  & \qw & \qw & \qw  \\
    & \gate{H} & \qw & \qw & \qw & \control \qw  & \qw & \gate{H} & \qw
    }
    }
    \caption{Local unitary equivalent circuit of $\ket{C_5}$ state that produces state $\ket{\psi_5}$}
    \label{fig:2}
\end{figure}

\begin{equation}\label{eq:1}
\begin{aligned}
    \ket{\psi_5}=L.U.\ket{C_5} = \frac{1}{2}(\ket{00000}+\ket{00111}+\ket{11100}+\ket{11011}).
\end{aligned}
\end{equation}

\begin{figure}[!hbt]
    \centerline{
    \Qcircuit @C=4em @R=.8em{
    & \lstick{\ket{0}} & \gate{H} & \ctrl{1} & \qw & \qw & \\
    & \lstick{\ket{0}} & \qw & \targ & \ctrl{1} & \qw & \\
    & \lstick{\ket{0}} & \gate{H} & \ctrl{1} & \targ & \qw & \rstick{\ket{\psi_5}} \\
    & \lstick{\ket{0}} & \qw & \targ & \ctrl{1} & \qw  &\\
    & \lstick{\ket{0}} & \qw & \qw & \targ & \qw 
    \gategroup{1}{6}{5}{6}{0.7em}{\}}
    }
    }
    \caption{Circuit diagram for $\ket{\psi_5}$ state}
    \label{fig:3}
\end{figure}

Here we have used the state $\ket{0}^{\otimes5}$ as input. Using 32 computational basis states as input to this circuit, we can obtain 32 mutually orthogonal five-qubit states, as provided in Table~\ref{table:4}. It can be easily seen that these 32 mutually orthogonal states are also LU equivalent of of five-qubit cluster state $\ket{C_5}$. In our proposed protocol, we use the state $\ket{\psi_5}$ for communication.
 
\subsection{Non Destructive Discrimination (NDD)}\label{NDD}
Non-destructive discrimination (NDD) is a measurement-based  approach which helps in discriminating orthogonal quantum states by measuring them without destroying the entanglement of the system. NDD has been implemented on Bell states in \cite{Samal_2010}.

\begin{figure}[!hbt]
    \centering
    \includegraphics[width=0.6\linewidth]{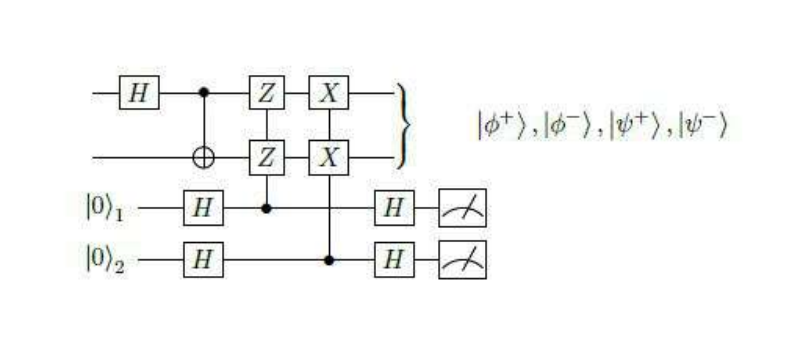}
    \caption{NDD of Bell States}
    \label{fig:4}
\end{figure}

 We briefly review NDD of Bell states. The four Bell states are, 

 \begin{align*}
      &\ket{ \phi^{\pm}} = \frac{1}{\sqrt{2}}(\ket{00} \pm \ket{11}), \text{ and } \ket{\psi ^{\pm}} = \frac{1}{\sqrt{2}}(\ket{01} \pm \ket{10}).
 \end{align*}

We consider the operators $X_1 X_2$ and $Z_1 Z_2$, and apply them on the four  Bell states.
\begin{equation}
    \label{eq:2}
    \begin{aligned}
        & X_1 X_2 \ket{\phi^+}  = \ket{\phi^+}, X_1 X_2 \ket{\phi^-} = -\ket{\phi^-},\\
        & X_1 X_2 \ket{\psi^+} = \ket{\psi^+}, X_1 X_2 \ket{\psi^+} = -\ket{\psi^+}.
    \end{aligned}
\end{equation}
\begin{equation}
    \label{eq:3}
    \begin{aligned}
        & Z_1Z_2 \ket{\phi^+} = \ket{\phi^+} , Z_1Z_2 \ket{\phi^-} = \ket{\phi^-},\\
        & Z_1Z_2 \ket{\psi^+} = -\ket{\psi^+}, Z_1Z_2 \ket{\psi^-} = -\ket{\psi^-}.
    \end{aligned}
\end{equation}

From Eq.s~\ref{eq:2}, \ref{eq:3}, we observe that Bell states are eigenvectors for the operators $X_1 X_2$ and $Z_1 Z_2.$
$X_1 X_2$ projects $\ket{\phi^+}$ ,$\ket{\psi^+}$ onto (+1) eigenspace and $\ket{\phi^-}$, $\ket{\psi^-}$ onto (-1) eigenspace.
Similarly, $Z_1 Z_2$ projects $\ket{\phi^+}$, $\ket{\phi^-}$ onto (+1) eigenspace and $\ket{\psi^+}$, $\ket{\psi^-}$ onto (-1) eigenspace. Fig.~\ref{fig:4} represents circuit diagram for NDD of Bell states using syndrome measurements. Ancillae qubits $\ket{0}_1$ and $\ket{0}_2$ are introduced to perform syndrome measurements of operators $X_1 X_2$ and $Z_1 Z_2$ respectively.

\begin{table}[!htb]
    \centering
    \begin{tabular}{|c|c|c|}
    \hline
    Bell State & Measurement of ancilla 1 & Measurement of ancilla 2\\
    \hline
      $\ket{\phi^+}$  & 0 & 0  \\
    $\ket{\phi^-}$ & 0 & 1 \\
    $\ket{\psi^+}$ & 1 & 0 \\
    $\ket{\psi^-}$ & 1 & 1\\
    \hline
    \end{tabular}
    \caption{Ancilla measurements for Non-Destructive Discrimination of Bell states.}
    \label{table:11}
\end{table}

If the state belongs to (+1) eigenspace, ancilla is measured '0', and if the state belongs to the (-1) eigenspace, ancilla is measured '1'. This provides a unique way to discriminate all four Bell states without destroying their entanglement. As an example, it can be easily seen that if the ancillae qubits are measured in state $\ket{00}$, the corresponding Bell state is $\ket{\phi^+}$. All the ancillae measurements corresponding to four Bell states has been shown in Table \ref{table:11}.

\begin{figure}[!hbt]
    \centering
    \scalebox{0.85}{ 
    \Qcircuit @C=1.6em @R=.3em{
     & \qw  & \qw & \gate{Z} \qwx[1] & \gate{X}  \qwx[1] & \qw & \qw & \qw & \qw & \qw & \qw \\
    & \qw  & \ctrl{1} & \gate{Z} & \gate{X} & \qw & \qw & \qw & \ctrl{1} & \qw & \qw \\
    \lstick{\ket{\psi_5}} & \qw & \targ  & \qw & \qw & \gate{Z} \qwx[1] & \gate{X} \qwx[1]  & \qw & \targ & \qw & \qw & \rstick{\ket{\psi_5}}  \\
    & \qw  & \ctrl{1} & \qw & \qw & \gate{Z} & \gate{X}  & \qw & \ctrl{1} & \qw & \qw  \\
    & \qw & \targ  & \qw & \qw & \qw & \qw & \gate{Z}  & \targ & \qw & \qw \\
    & \lstick{\ket{0}}  & \gate{H} & \ctrl{-4} & \qw & \qw & \qw & \qw & \gate{H} & \meter \\
    & \lstick{\ket{0}}  & \gate{H} & \qw & \ctrl{-5} & \qw & \qw & \qw & \gate{H} & \meter \\
    & \lstick{\ket{0}}  & \gate{H} & \qw & \qw & \ctrl{-4} & \qw & \qw & \gate{H} & \meter \\
    & \lstick{\ket{0}}  & \gate{H} & \qw & \qw & \qw & \ctrl{-5} & \qw & \gate{H} & \meter \\
    & \lstick{\ket{0}}  & \gate{H} & \qw & \qw & \qw & \qw & \ctrl{-5} & \gate{H} & \meter 
    \gategroup{1}{1}{5}{1}{0.7em}{\{}
    \gategroup{1}{11}{5}{11}{0.7em}{\}}
    }
    }
    \caption{NDD of $\ket{\psi_5}$ state }
    \label{fig:5}
\end{figure}

We now propose a scheme for performing the non-destructive discrimination of state $\ket{\psi_5}$ which utilizes the NDD of Bell states. First we use unitary transformations to convert the five-qubit (LU equivalent) cluster state into a state which is a product of Bell states on $2$ consecutive subsystems of $2$ consecutive qubits, while the last qubit is left in a separable single qubit state. Next, we apply the NDD on each of the 3 subsystems as described above, and later apply inverse of the  unitary operations to restore the entangled state $\ket{\psi_5}$. 

\begin{table}[!htb]
    \centering
    
    \resizebox{15cm}{!}{
    \begin{tabular}{|c|c|c|c|}
    \hline
     Classical Messages & Encoding Operations &States ($\Omega$) & Ancillae Outcome\\
    \hline
    00000 & $I_1 \otimes I_3 \otimes I_5$ & $\ket{\psi_5}^0= \frac{1}{2} ( \ket{00000}+ \ket{00111} + \ket{11100} + \ket{11011})$ & 00000\\
    00010 & $ Z_1 \otimes I_3 \otimes I_5$ & $\ket{\psi_5}^1= \frac{1}{2}(\ket{00000} + \ket{00111} - \ket{11100} -\ket{11011})$ & 01000\\
    10000 & $X_1 \otimes I_3 \otimes I_5 $ & $\ket{\psi_5}^2= \frac{1}{2} ( \ket{10000} + \ket{01100} + \ket{01011} + \ket{10111})$ & 10000\\
    00011 & $ Z_1 \otimes Z_3 \otimes I_5$ & $\ket{\psi_5}^3= \frac{1}{2} ( \ket{00000}+ \ket{11100}-\ket{00111}-\ket{11011})$ & 00010\\
    01000 & $ I_1 \otimes X_3 \otimes I_5$ & $\ket{\psi_5}^4 = \frac{1}{2}(\ket{11000}+ \ket{00100} + \ket{00011}+ \ket{11111})$ & 00100\\
    00100 & $ I_1 \otimes I_3 \otimes X_5$ & $\ket{\psi_5}^5= \frac{1}{2}( \ket{00001} + \ket{00110} + \ket{11010} + \ket{11101}$ & 00001\\
    10010 & $ XZ_1 \otimes I_3 \otimes I_5$ & $\ket{\psi_5}^6 = \frac{1}{2} (\ket{01100} + \ket{01011} - \ket{10111} - \ket{10000}$ & 11000\\
    00001 & $ I_1 \otimes Z_3 \otimes I_5$ & $\ket{\psi_5}^7 = \frac{1}{2} ( \ket{00000}- \ket{00111}-\ket{11100}+ \ket{11011})$ & 01010\\
    01010 & $Z_1 \otimes X_3 \otimes I_5$ & $\ket{\psi_5}^8 = \frac{1}{2}( \ket{00100} + \ket{00011} - \ket{11000}- \ket{11111})$ & 01100\\
    00110 & $Z_1 \otimes I_3 \otimes X_5$ & $\ket{\psi_5}^9 = \frac{1}{2} (\ket{00001}+ \ket{00110}-\ket{11101}-\ket{11010})$ & 01001\\
    10011 & $XZ_1 \otimes Z_3 \otimes I_5$ & $\ket{\psi_5}^{10}= \frac{1}{2}( \ket{10000} + \ket{01100}- \ket{01011} - \ket{10111})$ &  10010\\
    11000 & $X_1 \otimes X_3 \otimes I_5$ & $\ket{\psi_5}^{11}= \frac{1}{2}( \ket{01000}+ \ket{10100} + \ket{10011} +\ket{01111})$ & 10100\\
    10100 & $X_1 \otimes I_3 \otimes X_5$ & $\ket{\psi_5}^{12} = \frac{1}{2}(\ket{01010} + \ket{10110}+ \ket{10001} +\ket{01101})$ & 10001\\
    01011 & $Z_1 \otimes XZ_3 \otimes I_5$ & $\ket{\psi_5}^{13}= \frac{1}{2}(\ket{00011}-\ket{11000}-\ket{00100}+\ket{11111})$ & 00110\\
    00111 & $Z_1 \otimes Z_3 \otimes X_5$ & $\ket{\psi_5}^{14}= \frac{1}{2}(\ket{00001}+ \ket{11101}- \ket{00110}-\ket{11010})$ & 00011\\
    01100 & $I_1 \otimes X_3 \otimes X_5$ & $\ket{\psi_5}^{15}= \frac{1}{2}(\ket{00010}+ \ket{11110}+\ket{11001}+ \ket{00101})$ & 00101\\
    10001 & $X_1 \otimes Z_3 \otimes I_5$ & $\ket{\psi_5}^{16}= \frac{1}{2}(\ket{01100} + \ket{10111}-\ket{01011}-\ket{10000})$ & 11010\\
    11010 & $XZ_1 \otimes X_3 \otimes I_5$ & $\ket{\psi_5}^{17}= \frac{1}{2}(\ket{01000}+ \ket{01111}-\ket{10011}-\ket{10100})$ & 11100\\
    10110 & $XZ_1 \otimes I_3 \otimes X_5$ & $\ket{\psi_5}^{18}= \frac{1}{2}(\ket{01010}-\ket{10110}-\ket{10001}+\ket{01101})$ & 11001\\
    01001 & $I_1 \otimes XZ_3 \otimes I_5$ &$\ket{\psi_5}^{19} = \frac{1}{2}(\ket{11000}- \ket{00100}+\ket{00011}-\ket{11111})$ & 01110\\
    00101 & $I_1 \otimes Z_3 \otimes X_5$ & $\ket{\psi_5}^{20}= \frac{1}{2}(\ket{11010}-\ket{00110}+\ket{00001}-\ket{11101})$ & 01011\\
    01110 & $Z_1 \otimes X_3 \otimes X_5$ & $\ket{\psi_5}^{21}= \frac{1}{2}(\ket{00010}-\ket{11110}-\ket{11001}+\ket{00101})$ & 01101\\
    11011 & $ XZ_1 \otimes XZ_3 \otimes I_5$ &$\ket{\psi_5}^{22}=\frac{1}{2}(\ket{01111}+\ket{10011}-\ket{10100}-\ket{01000})$ &  10110\\
    10111 & $XZ_1 \otimes Z_3 \otimes X_5$ &$\ket{\psi_5}^{23}= \frac{1}{2}(\ket{10001}+\ket{01101}-\ket{10110}-\ket{01010})$ & 10011\\
    11100 & $X_1 \otimes X_3 \otimes X_5$ & $\ket{\psi_5}^{24}= \frac{1}{2}(\ket{10010}+\ket{01110}+\ket{01001}+\ket{10101})$ & 10101\\
    01111 & $Z_1 \otimes XZ_3 \otimes X_5$ & $\ket{\psi_5}^{25}= \frac{1}{2}(\ket{00010}+\ket{11110}-\ket{11001}-\ket{00101})$ & 00111\\
    11001 & $X_1 \otimes XZ_3 \otimes I_5$ & $\ket{\psi_5}^{26}= \frac{1}{2}(\ket{10100}-\ket{10011}+\ket{01111}-\ket{01000})$ & 11110\\
    10101 & $X_1 \otimes Z_3 \otimes X_5$ & $\ket{\psi_5}^{27}= \frac{1}{2}(\ket{10110}+\ket{01101}-\ket{10001}-\ket{01010})$ & 11011\\
    11110 & $XZ_1 \otimes X_3 \otimes X_5$ & $\ket{\psi_5}^{28}= \frac{1}{2}(\ket{01001}+\ket{01110}-\ket{10101}-\ket{10010})$ & 11101\\
    01101 & $I_1 \otimes XZ_3 \otimes X_5$ & $\ket{\psi_5}^{29}= \frac{1}{2}(\ket{00010}-\ket{11110} + \ket{11001}-\ket{00101})$ & 01111\\
    11111 & $XZ_1 \otimes XZ_3 \otimes X_5$ & $\ket{\psi_5}^{30}= \frac{1}{2}(\ket{10010} + \ket{01110}-\ket{01001}-\ket{10101})$ & 10111\\
    11101 & $X_1 \otimes XZ_3 \otimes X_5$ & $\ket{\psi_5}^{31}= \frac{1}{2}(\ket{01110}+\ket{10101}-\ket{01001}-\ket{10010})$ & 11111\\
    \hline
    \end{tabular}}
    \caption{Encoding Unitary Operations for classical messages, and their corresponding outcome states and ancillae measurements required for NDD of that particular outcome.}
    \label{table:4}
\end{table}

The circuit diagram to perform the NDD of $\ket{\psi_5}$ state is provided in Fig.~\ref{fig:5}. We use unitary operations $C_2X_3$ and $C_4X_5$ to convert $\ket{\psi_5}$ into $\ket{\phi^+}_{12} \otimes \ket{\phi^+}_{34} \otimes \ket{0}_{5}$. Consulting Table~\ref{table:4}, one can easily see that in this particular case, the ancillae will be in state $\ket{00000}$. This circuit can be used to identify each of the $32$ mutually orthogonal states achieved from by changing the input of $\ket{\psi_5}$. The ancillae outcome corresponding to the individual states are provided in Table~\ref{table:4}. As an example, for state $\ket{\psi_5}^{11}=\frac{1}{2}( \ket{01000}+ \ket{10100} + \ket{10011} +\ket{01111})$, the ancillae outcome is $\ket{10100}$.

\begin{figure}[!hbt]
    \centering
    \includegraphics[width=0.8\linewidth]{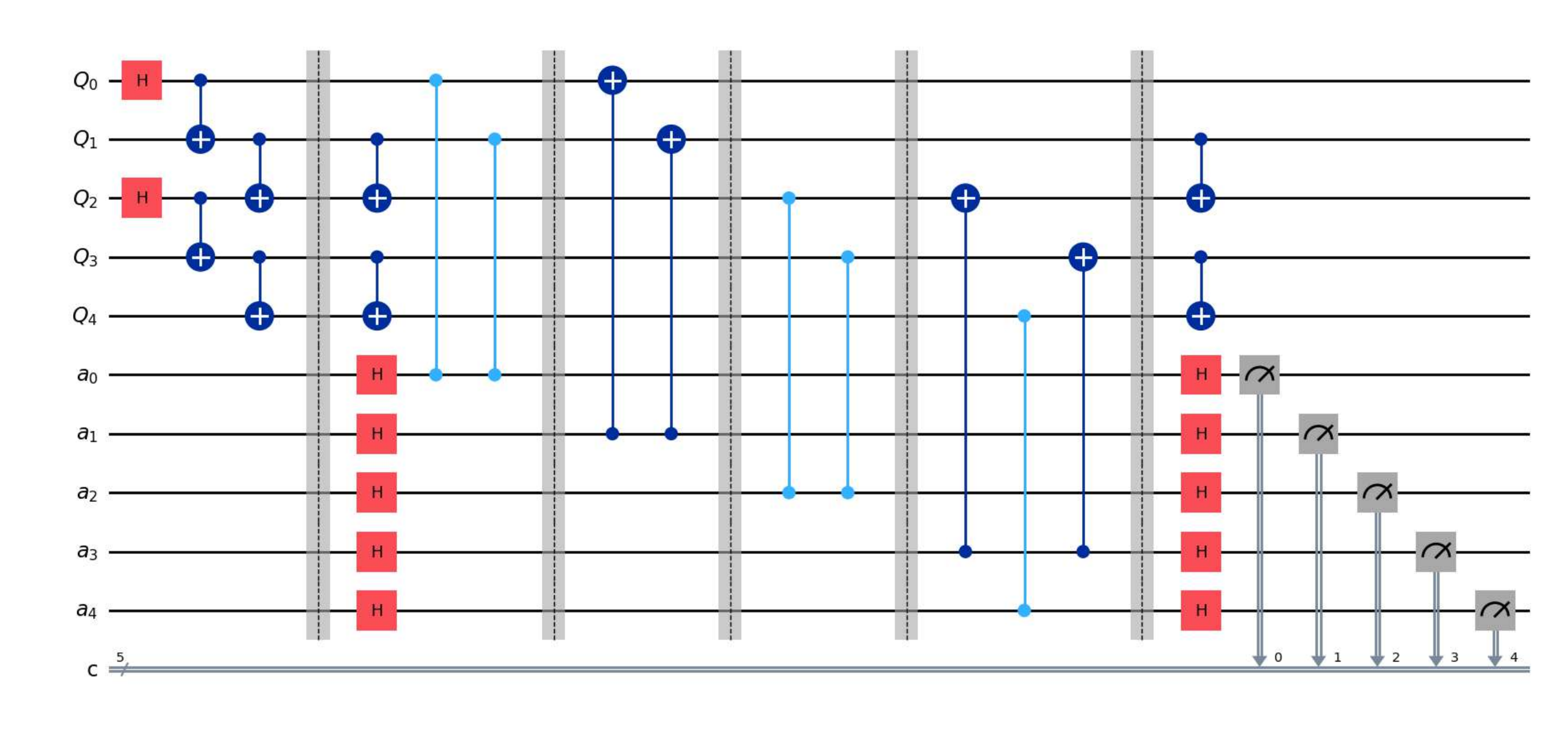}
    \caption{NDD implementation in qiskit}
    \label{fig:6}
\end{figure}

 We have implemented the NDD circuit in  IBM qiskit \cite{javadiabhari2024quantumcomputingqiskit} using the Aer Simulator therein. The corresponding circuit is provided in Fig.~\ref{fig:6}. As can be seen from Fig.~\ref{fig:6}, here we have taken $\ket{\psi_5^{11}} = \frac{1}{2}(\ket{01000}+ \ket{01111}+\ket{10100}+\ket{10011})$ as the input to the NDD protocol. 
 
\begin{figure}[!hbt]
    \centering
    \includegraphics[width=0.75\linewidth]{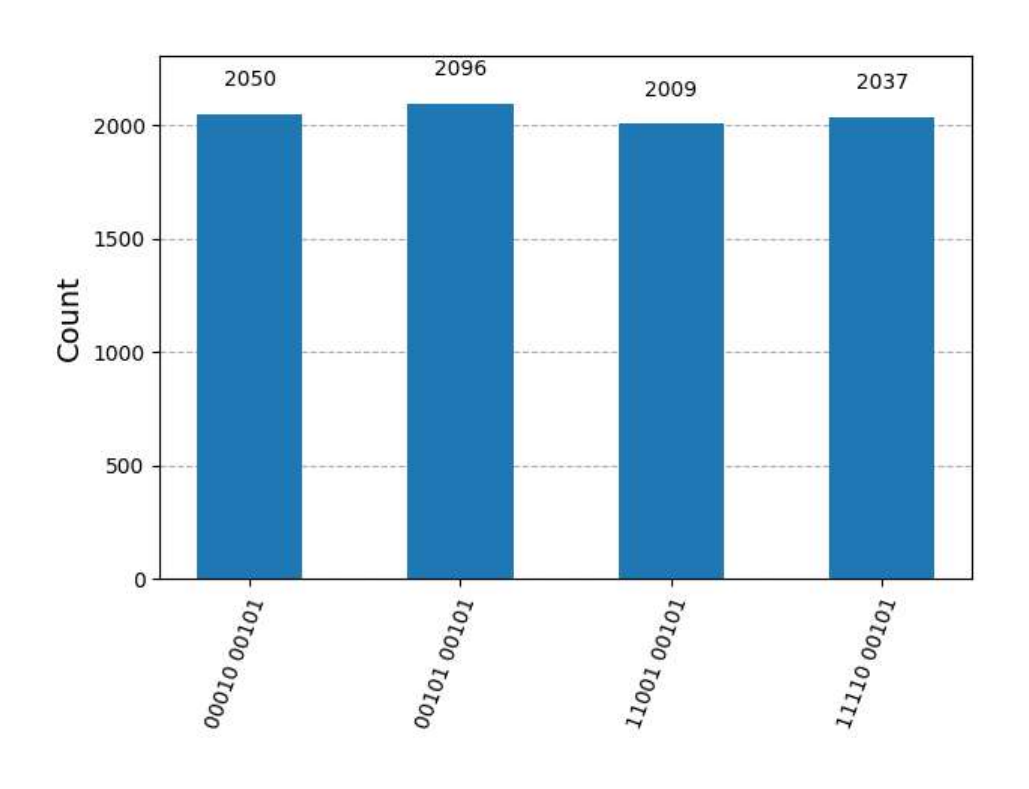}
    \caption{Measurement outcome of NDD of $\ket{\psi_5^{11}}$ state.}
    \label{fig:7}
\end{figure}
Fig.~\ref{fig:7} shows the measurement outcome of the circuit in Fig.~\ref{fig:6} with $8192$ shots. From the histogram plot of the output, we can see that, the $5$ qubits corresponding to the ancillae qubits are in state $\ket{10100}$ indicating the syndrome of the state $\ket{\psi_{5}}^{11}$, whereas the remaining $5$ qubits are in an almost equal superposition of the basis states $\ket{01000}, \ket{01111}, \ket{10100}, \ket{10011}$, mimicking state $\ket{\psi_{5}}^{11}$. Experimental realization of NDD for different states have been carried out in \cite{article, SISODIA20173860}.

\section{Encoding Scheme}\label{sec:3}
The quantum dialogue scheme requires encoding and sharing of a message. In our protocol, Alice and Bob use the same set of unitary operations to encode the message. The encoding process is similar to a quantum dense coding protocol, however, we require that these unitary operations follow a group theoretic structure to ensure the NDD based scheme works.

We assume $U_A$ and $U_B$ are the unitary operations required to encode  Alice's and Bob's messages, respectively. First, Alice encodes her message on state $\ket{\psi_5}$ and sends it to Bob. Since in our case, the quantum state is preserved through NDD, Bob encodes his message on the state he received, and sends it back to Alice. The corresponding quantum operations can be found as, 
$$ \ket{\psi_5} \rightarrow U_A \ket{\psi_5} \rightarrow U_B U_A \ket{\psi_5}.$$

We now list two conditions that need to be fulfilled by the set of encoding unitaries for our scheme to be successful. 

\begin{itemize}
    \item Performing $U_A$ on $\ket{\psi_5}$ transforms it to one of the $32$ mutually orthogonal states provided in Table~\ref{table:4}, say, $\ket{\psi_5}^{L_1}$. Similarly, when $U_B$ is performed on $\ket{\psi_5}^{L_1}$, it is yet again transformed into another one of these $32$ mutually orthogonal states, say, $\ket{\psi_5}^{L_2}$.
    
    \item $U_B^jU_A^i$, $\forall i, j$ individually maps $\ket{\psi_5}$ to a unique state among the set of $31$ quantum states that are mutually orthogonal to $\ket{\psi_5}$. 
\end{itemize}

These constructional constraints on the set of encoding unitary operations used by Alice and Bob $\{ U_{\{A/B\}}^i \}$ ensure that  
they form a group $\mathcal{U}_{\{A, B\}}$, such that all three of  $ U_A, U_B, \text{ and } U_BU_A \in \mathcal{U}_{\{A, B\}}$ are such that they transform one of the $32$ mutually orthogonal states provided in Table~\ref{table:4} to another. 
The formation of the group $\mathcal{U}_{\{A, B\}}$ also ensures that when the state $U_BU_A\ket{\psi_5}$ is reverted back to Alice, she can perform NDD to decipher Bob's message. If the unitary operations $U_A, U_B$ do not form a group, then the operation $U_B U_A$ might not transform $\ket{\psi_5}$ into one of the 32 orthogonal states. It might also happen that for different set of unitary operations, we may get same state. As an example, say Bob uses unitaries $U_B^1$ and $U_B^2$ to encode two different messages $m_1$ and $m_2$ respectively. However, if in this case, $U_B^1U_A\ket{\psi_5}= U_B^2U_A\ket{\psi_5}$, Alice will not be able to discriminate Bob's actual messages, as the outcome of her NDD protocol will be same for both $m_1$ and $m_2$, leading to the failure of the dialogue protocol. 

In this protocol, for each encoding message, we need different unitary operations. Following the above discussion, each unitary operation yields a unique quantum state. As the NDD protocol can only discriminate between $32$ mutually orthogonal quantum states with a five-qubit state, the order of the group of encoding unitary operations $\mathcal{U}_{\{A, B\}}$ is $32$. However, encoding is done only on three-qubits, i.e., each $U_e \in \mathcal{U}_{\{A, B\}}$ acts on a Hilbert space of dimension $8$.

Since $G= \lbrace I, X, Z, XZ \rbrace$ is a group of order 4
and $S_G = \lbrace I , X \rbrace$ is a subgroup of G of order 2, we define the encoding unitary operations $U_e \in \mathcal{U}_{\{A, B\}}$ as,  $$ U_e = U_G \otimes U_G \otimes U_{S_G}, $$ where, $U_G$ is an element of $G$ and $U_{S_G}$ is an element of $S_G$. As $G$ and $S_G$ are groups, it can be easily seen that $\mathcal{U}_{\{A, B\}}$ forms a group as well. We provide a proof of the same in Appendix~\ref{Appen}. All $U_e \in \mathcal{U}_{\{A, B\}}$ are listed in Table~\ref{table:4}.

We now discuss the encoding scheme of the classical messages using group $\mathcal{U}_{\{A, B\}}$. In the beginning of the protocol, Alice holds $3$ of the $5$ qubits in $\ket{\psi_5}$, i.e., qubits $1, 3, 5$, and encodes her message on these three qubits only. As per the construction of  $\mathcal{U}_{\{A, B\}}$, Alice has $32$ different combinations of unitary gates to encode a five bit classical message on the state $\ket{\psi_5}$. She can encode a total of $32$ classical messages in this way. The encoding follows this simple rule:
For classical message $m= "m_1m_2m_3m_4m_5"$, $m_i \in \lbrace 0,1\rbrace, i \in \{1, 2, 3, 4, 5\}$, the encoding unitary is, 
\begin{align}\label{encoder}
    {U_e}^m= X^{m_1}Z^{m_4} \otimes X^{m_2}Z^{m_5} \otimes X^{m_3}
\end{align}

For example, if Alice wants to send '10011' to Bob, then the unitary operation for encoding will be $XZ_1 \otimes Z_3 \otimes I_5$ and the resulting state would be $\ket{\psi_5^{10}}$. Table \ref{table:4} shows all the cases of the ancillae measurements corresponding to all $32$ orthogonal  states. We provide a proof for bijection between classical message and ancilla outcome in Appendix ~\ref{Appen}.

\section{Protocol}
\label{sec:4}
We now present our protocol for quantum dialogue scheme. In the proposed protocol, $\ket{\psi_5}$ is used for communication. Both parties can send classical messages of at most five bits, and information is encoded on the first, third and fifth qubits of the state in both cases. When information is encoded and transmitted, at the receiver's end, NDD is performed to decipher the transmitted state. As the group theoretic architecture of encoding unitaries ensures uniqueness of the product of every pair of group elements, the receiver is able to decipher the corresponding unitary operation from the NDD outcome, and the message is decoded using Eq.~\ref{encoder}. 

An important highlight of the protocol is that, in absence of an Eavesdropper, no classical communication is required for decoding the message. 

 The protocol is as follows:

\begin{enumerate}
    \item We assume, Alice wants to send a classical message of bit-length $N_c= 5N$. She prepares N copies of $\ket{\psi_5}$ state. Then, Alice breaks her $N_c$ bit message into $N$ $5-bit$ messages, and encodes her them in the first, third and fifth qubits of the $N$ state using the predefined encoding scheme. For the rest of this article, we consider the encoding scheme to be as given in Eq.~\ref{encoder}. 
    
    \item Next, Alice chooses a single qubit located at the same position from each of $N$ states, and forms five ordered sequences, each consisting $N$ qubits. These sequences are 
    $$\{S_i = \lbrace q_i^1, q_i^2, \dots q_i^N \rbrace\}_{i=1}^5,$$
    where $q_i$ is the $i^{th}$ qubit from $\ket{\psi_5}$ and subscripts denotes the sequential ordering of $\ket{\psi_5}.$
    
    \item To prevent eavesdropping, Alice prepares $m=5N$ decoy qubits. The decoy qubits are randomly prepared in one of the four states $\lbrace \ket{0}, \ket{1}, \ket{+}, \ket{-} \rbrace$ where $\ket{+}=\frac{\ket{0} + \ket{1}}{2}$ and $\ket{-}= \frac{\ket{0} - \ket{1}}{2}.$ She then inserts $N$ decoy qubits randomly in each of the five sequences, thus creating new sequences $\{S_1',S_2',S_3',S_4',S_5'\}$ each containing $2N$ qubits.
    
    \item  Alice sends these sequences one by one in reverse order to Bob. First she sends the fifth sequence, i.e., $S_5'$ to Bob. After confirming that Bob has received the sequence, Alice announces the position of decoy qubits. Bob measures the corresponding qubits in sequence $S_5'$ by using X-basis or Z-basis at random where $X= \lbrace \ket{0}, \ket{1} \rbrace$ and $Z= \lbrace \ket{+}, \ket{-} \rbrace.$ After measurement Bob informs his measurement outcome and bases to Alice who computes the error rate. If sufficiently few errors are found, they go to next step otherwise they truncate the protocol. 
    
    \item If no eavesdropping is found in the previous step, Alice send sequence $S_4'$ to Bob and checks for eavesdropping. If eavesdropping is found, the protocol is truncated other wise, Alice send the remaining three sequences $S_3'$, $S_2'$, $S_1'$ to Bob in similar manner.
    
    \item Now Bob performs non destructive discrimination on the unmeasured part of the transmitted state(s). As explained in Sec.~\ref{NDD}, the output of Bob's measurement with input $\ket{\psi_5^i}$ will be, $\ket{\psi_5^i}\ket{x_1x_2x_3x_4x_5}_i$ , where $\ket{\psi_5^i}$ is the transmitted state and $\ket{x_1 x_2 x_3 x_4 x_5}_i$ is the state of the ancilla, unique to $\ket{\psi_5^i}$. Additionally, it is one of the computational basis states. So, by measuring ancillae qubits, Bob deterministically determines the transmitted state and hence decodes Alice's message(s).
    
    \item Since Bob now has all the five-qubits, he can encode his classical message using the encoding principle Alice. He then follows steps 3-5 to send the state to Alice.
    
    \item In absence of any eavesdroppers, Alice performs non destructive discrimination to obtain classical message sent by Bob. Thus, the process continues and dialogue is established.
\end{enumerate}
\tikzstyle{process} = [rectangle, rounded corners,
minimum width= 1 cm, 
minimum height=2 cm, 
text centered, 
text width=3 cm, 
draw=black, 
fill=yellow!30]

\tikzstyle{process2} = [rectangle, rounded corners,
minimum width= 3 cm, 
minimum height=1 cm, 
text centered, 
text width=4 cm, 
draw=black, 
fill=yellow!30]

\tikzstyle{for} = [trapezium,
trapezium left angle=70,
trapezium right angle=110,
minimum width=0.5cm,
minimum height=1cm,
text width = 2cm,
text centered,
draw=black,
fill=blue!30]

\tikzstyle{decision} = [rectangle, 
minimum width=3cm,
minimum height=1cm, 
text centered, 
text width=3 cm,
draw=black,
fill=green!30]

\tikzstyle{decision2} = [diamond,
minimum width=0.3 cm,
minimum height=0.3 cm,
text centered,
text width = 2cm,
draw=black,
fill=purple!30]

\tikzstyle{stop} = [circle, 
minimum width=1cm, 
minimum height=0.5cm, 
text centered, 
text width=2cm,
draw=black, 
fill=red!60]

\tikzstyle{arrow} = [thick,->,>=stealth]
\begin{figure}
    \centering
    \label{fig:placeholder}
\resizebox{16cm}{!}{
\begin{tikzpicture}[node distance=2cm]
\node (start) [process] {Alice prepares $N$ copies of $\ket{\psi_5}$};
\node (step1) [process, right of= start, xshift=2cm] {Alice encodes her information and makes five ordered sequences  $S_i = \lbrace q_i^1, q_i^2, \dots q_i^N \rbrace $, i=1-5};
\node (step2) [process2, right of = step1, xshift=2.5cm] {Alice prepares m=5N decoy qubits randomly in $X$ or $Z$ bases and randomly inserts in five sequences $S_i$. Resulting sequences are $S_i'$};
\node (step3) [for, right of = step2, xshift= 2.5cm] {For i = 5 to 1, Alice sends sequence $S_i'$};
\node (step4) [decision, above of = step3, yshift=2cm] {Alice announces the position of decoy qubits};
\node (step5) [decision, above of = step4, yshift=2cm] {Bob measures the decoy qubits randomly in $X$ or $Z$ basis};
\node (step6) [decision, right of = step5, xshift= 2.5cm] {Bob announces his measurement outcome and bases to Alice};
\node (step7) [decision2, below of = step6, yshift=-1.8cm] {Is the error rate $< 25\% ?$ };
\node (step8) [stop, right of = step7, xshift= 2cm] {Abort the protocol};
\node (step9) [decision2, below of = step7, yshift=-2cm]{Has Bob recieved all the sequences?};
\node (step10) [process, right of = step9, xshift=2.5cm] {Bob performs NDD and decodes the message};

\draw [arrow] (start) -- (step1);
\draw [arrow] (step1) -- (step2);
\draw [arrow] (step2) -- (step3);
\draw [arrow] (step3) -- (step4);
\draw [arrow] (step4) -- (step5);
\draw [arrow] (step5) -- (step6);
\draw [arrow] (step6) -- (step7);
\draw [arrow] (step7) -- node[anchor=south] {No} (step8);
\draw [arrow] (step7) -- node[anchor=east] {Yes} (step9);
\draw [arrow] (step9) -- node[anchor=south] {Yes} (step10);
\draw [arrow] (step9) -- node[anchor=south] {No} (step3);
\end{tikzpicture}}
\caption{Flowchart for one-way communication between Alice and Bob.}
\end{figure}
\subsection{Examples}
\label{sec:4.1}
Let us understand the protocol using two examples. First under ideal condition, that is in absence of Eve. Second in presence of Eve. Consider the first case, absence of Eve. 
Let us assume, Alice's message is '10011'. As her message is consisted of just $5-$bits, Alice prepares only one five-qubit register in state $\ket{\psi_5}$. Following Eq.~\ref{encoder}, the encoding unitary corresponding to Alice's message is $XZ_1 \otimes Z_3 \otimes I_5$. She encodes her message and sends the state to Bob. For simplicity, we assume there is no eavesdropper present in this case. Upon receiving the state, Bob performs NDD, and finds the ancilla outcome to be '10010', which corresponds to state $\ket{\psi_5^{10}}$. Hence he deciphers the unitary operation, and decodes the message.\\

Now, Bob encodes his message '01110' using the prescribed unitary operation $Z_1 \otimes X_3 \otimes X_5$. He applies the unitary operations on the (intermediate state) $\ket{\psi_5^{10}}$ state, and converts it to $\ket{\psi_5^{31}}$.

Next, Alice performs NDD on state $\ket{\psi_5^{31}}$ as she receives it, and finds the ancilla outcome to be '11111'. This state which corresponds to $\ket{\psi_5^{31}}$. Since Alice knows the initial state as well as her encoding operation, she finds Bob's unitary operation and thus decodes the message.

\begin{figure}[!hbt]
    \centering
    \begin{minipage}[t]{0.48\textwidth}
        \centering
        \includegraphics[width=\linewidth]{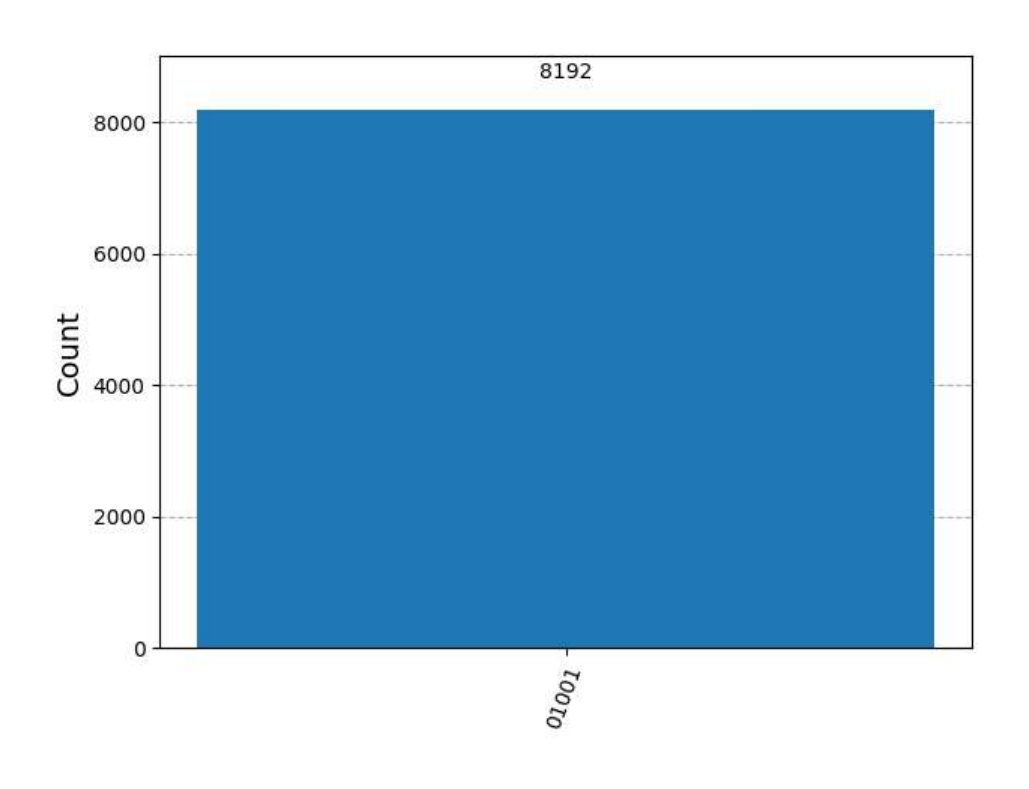}
        \caption{Outcome obtained when Bob performs NDD on the transmitted state. Transmitted state is $\ket{\psi_5^{10}}$.}
        \label{fig:8}
    \end{minipage}
    \hfill
    \begin{minipage}[t]{0.48\textwidth}
        \centering
        \includegraphics[width=\linewidth]{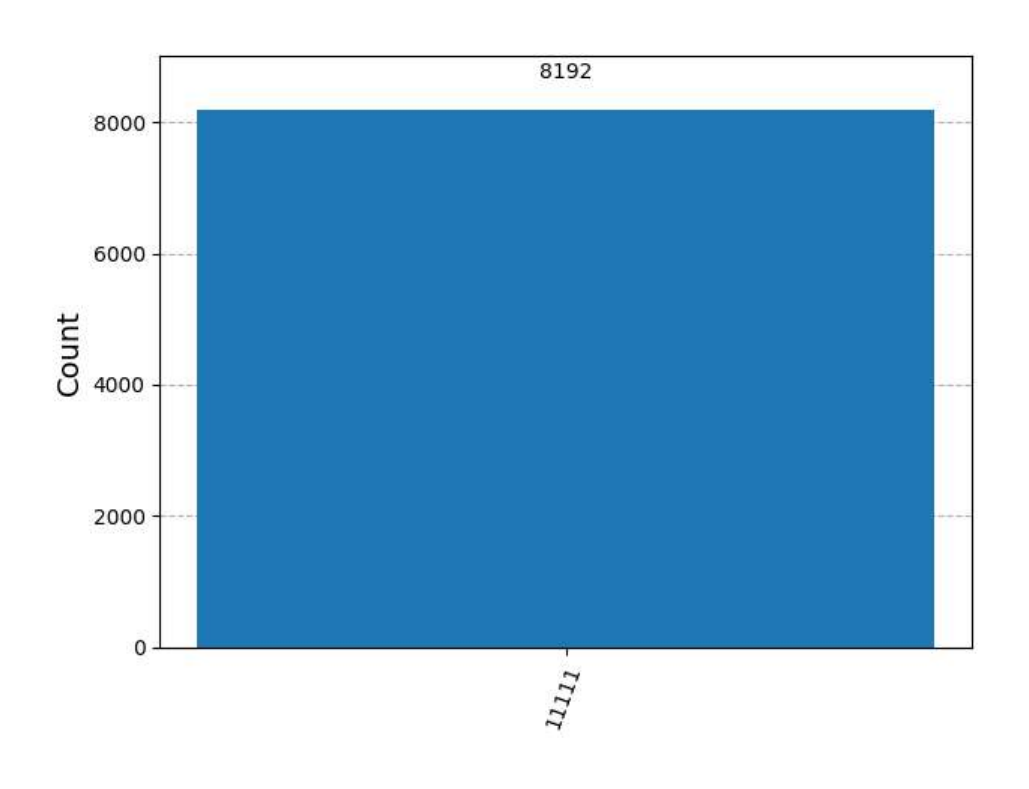}
        \caption{Outcome obtained when Alice performs NDD on the transmitted state. Transmitted state is $\ket{\psi_5^{31}}$.}
        \label{fig:9}
    \end{minipage}
\end{figure}

We have implemented this example on IBM qiskit using the ideal Aer Simulator, Fig.s~\ref{fig:8}, \ref{fig:9} shows the ideal ancilla outcomes for Bob and Alice respectively.

\begin{figure}[!hbt]
    \centering
    \begin{minipage}[t]{0.48\textwidth}
        \centering
        \includegraphics[width=\linewidth]{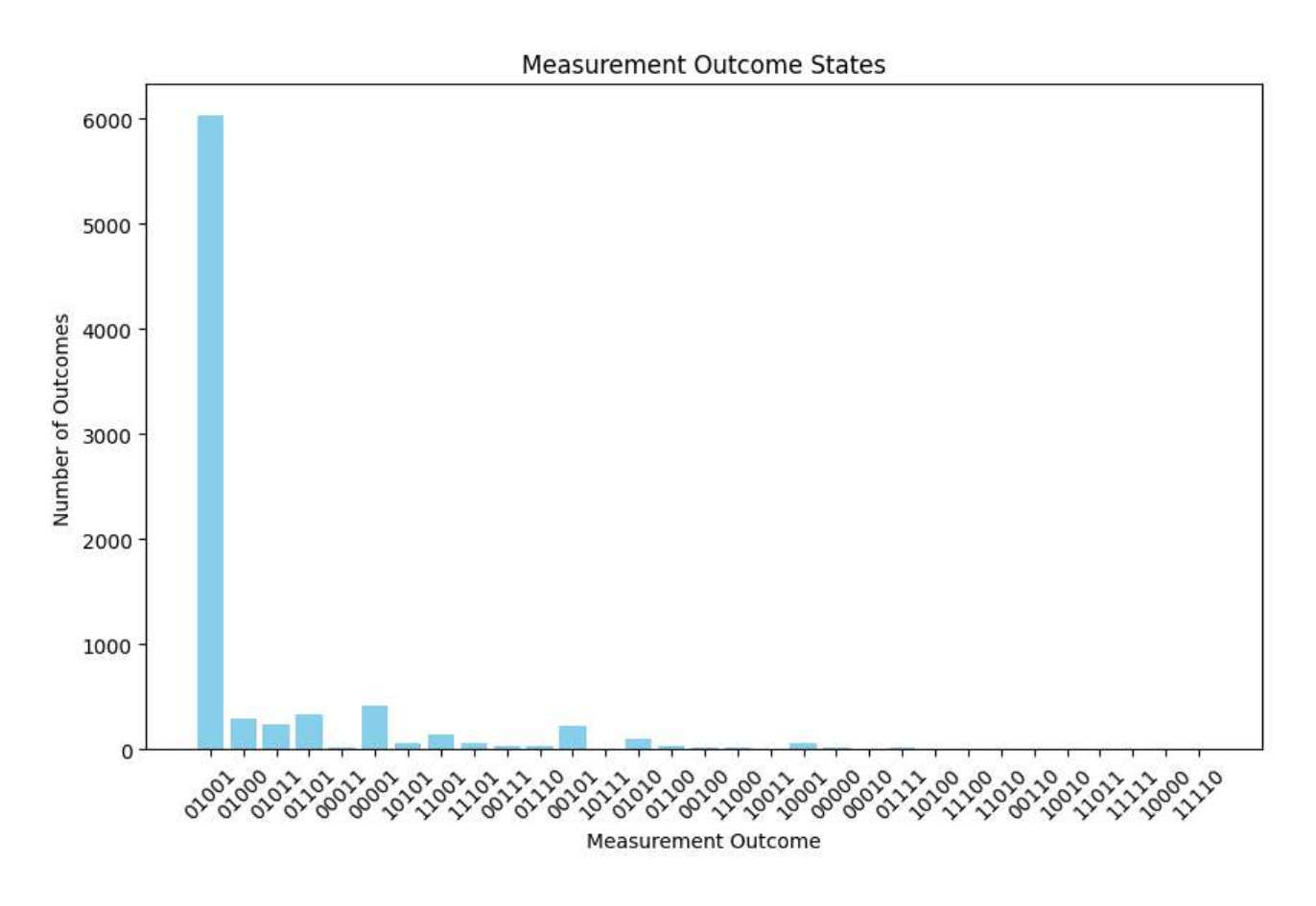}
        \caption{Ancilla outcome obtained when Bob performs NDD on the transmitted state $\ket{\psi_5^{10}}$. Backend used is \texttt{IBM\_Torino}.}
        \label{fig:10}
    \end{minipage}
    \hfill
    \begin{minipage}[t]{0.48\textwidth}
        \centering
        \includegraphics[width=\linewidth]{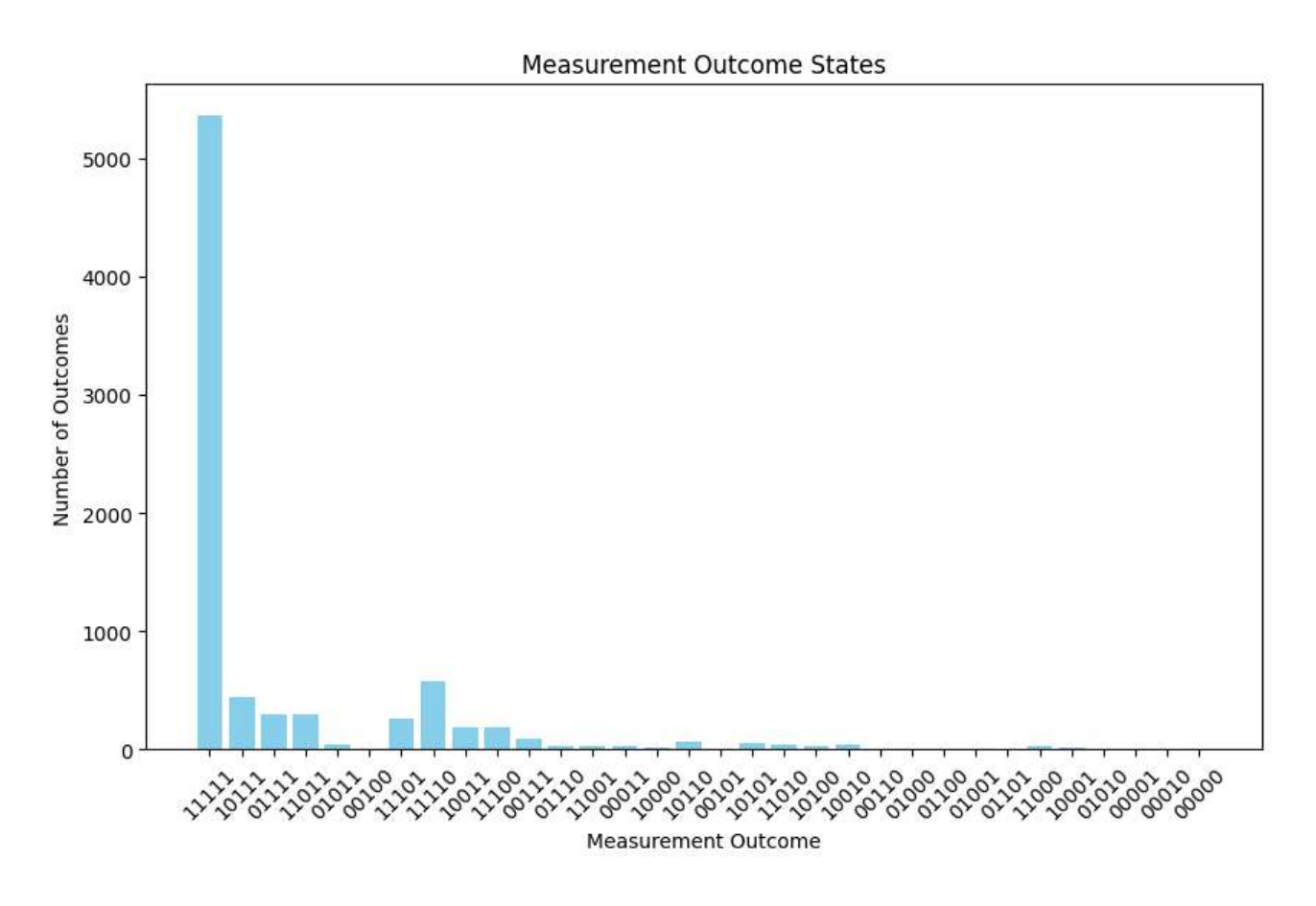}
        \caption{Ancilla outcome obtained when Alice performs NDD on the transmitted state $\ket{\psi_5^{31}}$. Backend used is \texttt{IBM\_Torino}.}
        \label{fig:11}
    \end{minipage}
\end{figure}

Further, we have implemented the quantum dialogue scheme on IBM's quantum hardware 'ibm\_torino'. Fig.~\ref{fig:10}  shows the ancilla outcomes when Bob performs NDD as found in the real device, without any noise mitigation or error correction with $8192$ shots. It can be seen that the probability of finding the actual outcome '10010' surpasses the other possible outcomes. Similarly, Fig.~\ref{fig:11} shows ancilla outcomes when Alice performs NDD. We observe that '11111' is found with highest probability, as is expected from the protocol. Despite using noisy real quantum hardware without any noise mitigation or error correction, the deterministic nature of the protocol outcome leads to a successful outcome, which is significantly rare. However, this experimental result is provided in this manuscript as a proof of concept. Further detailed studies are required to determine the applicability of the protocol on real quantum hardware.\\
Now let us consider the second case where Eve is present in between. Suppose Alice wants to send the message '1010011010'. She will divide the message in two parts '10100' and '11010' respectively. She will prepare two copies of $\ket{\psi_5}$. On the first copy she will encode '10100' and '11010' on the second copy using the corresponding encoding operation. Using these two states, Alice will prepare five ordered sequences $S_i = \lbrace q_i^1,q_i^2 \rbrace $ for i= 1 to 5. Now Alice prepares 10 decoy qubits randomly in $X$ or $Z$ basis as $D= \lbrace \ket{0}, \ket{1}, \ket{+}, \ket{-}, \ket{0}, \ket{+}, \ket{1}, \ket{0}, \ket{-}, \ket{1} \rbrace$. Alice randomly inserts them in each of the five sequences to get new sequences $S_i'$.
$$ S_1'=\lbrace \ket{0}, q_1^1,q_1^2, \ket{1} \rbrace, \hspace{0.2cm} S_2' = \lbrace q_2^1, \ket{+}, \ket{-}, q_2^2 \rbrace, \hspace{0.2cm} S_3'= \lbrace \ket{0}, q_3^1, q_3^2, \ket{+} \rbrace$$
$$ S_4'= \lbrace \ket{1}, \ket{0}, q_4^1, q_4^2 \rbrace \hspace{0.2cm} S_5'= \lbrace q_5^1, q_5^2, \ket{-}, \ket{1} \rbrace.$$
\begin{table}[!hbt]
    \centering
    \resizebox{16cm}{!}{
    \begin{tabular}{|c|c|c|c|c|c|c|c|}
    \hline
    Sequence & Alice decoy state & Alice Basis & Eve basis & Eve measurement & Bob basis & Bob measurement & Mismatch (error)\\
    \hline
     \multirow{2}{2em}{$S_5$} &$\ket{1}$ & Z-basis & Z-basis & $\ket{1}$ & Z-basis & $\ket{1}$ & No \\
    & $\ket{+}$ & X-basis & X-basis & $\ket{+}$ & Z-basis & $\ket{0}$ or $\ket{1}$ & N/A \\
    \hline
    \multirow{2}{2em}{$S_4$} & $\ket{0}$ & Z-basis & X-basis & $\ket{+}$ or $\ket{-}$ & X-basis & $\ket{+}$ or $\ket{-}$ & N/A \\
     & $\ket{1}$ & Z-basis & Z-basis & $\ket{1}$ & Z-basis & $\ket{1}$ & No\\
    \hline
    \multirow{2}{2em}{$S_3$} & $\ket{+}$ & X-basis & X-basis & $\ket{+}$ & Z-basis & $\ket{0}$ or $\ket{1}$ & N/A\\
    & $\ket{0}$ & Z-basis & X-basis & $\ket{+}$ or $\ket{-}$ & Z-basis & $\ket{0}$ & Mismatch \\
    \hline
    \end{tabular}}
    \caption{Eavesdropping check}
    \label{table:30}
\end{table}
From table \ref{table:30} we observe that, when Alice's and Bob's basis does not match, we discard these situations. We only consider the situations when their basis matches. We observe that sequences $S_5$, $S_4$ are communicated without Eve getting detected. But while communicating sequence $S_3$, Eve gets detected and hence the protocol is aborted.

\section{Security Analysis}
\label{sec:5}
We now consider different Eavesdropping attacks on our protocol, and discuss the security of the protocol against them. 

\subsection{Passive Attack}
Information leakage is the difference between the total information sent between Alice and Bob and the minimum information required for Eve to extract the complete information. In a passive attack, Eve tries to get the secret message using the known information and without disturbing the communication. For one cycle of communication, Eve should at least know one of the unitary operations performed by Alice or Bob. Since there is bijection between encodings and the set of orthogonal states, the resulting 32 encoding states forms a complete , orthonormal basis for $2^5=32$-dimensional Hilbert space. As the message to be sent is random, the resulting encoded state could be any of the 32 orthogonal states. So, probability of any state $p_i= \frac{1}{32}$. As Eve does not know the unitary operations performed, Eve can posses a maximally mixed reduced density matrix over 32-dimensional Hilbert space. 
$$ \rho_{Eve}= \sum_{i=1}^{32} \rho_i \ket{\psi_5^i} \bra{\psi_5^i} = \frac{1}{32} \sum_{i=1}^{32} \ket{\psi_5^i} \bra{\psi_5^i} = \frac{I}{32}$$.
The information required for Eve to extract the complete information is given by von Neumann entropy $$ S(\rho_{Eve}) = -Tr(\rho_{Eve} \log_2 \rho_{Eve}) = -\sum_{i=1}^{32} \rho_i \log_2 \rho_i = -32 \times \frac{1}{32} \log_2 \frac{1}{32}= 5-bits$$ 
Thus the amount of information Eve requires to know to decipher the message is the total information sent between Alice and Bob. Hence no-information leakage happens in the proposed protocol. 

\subsection{Man-in-the-middle attack}
The Man-in-the-middle (MITM) attack is a scenario in which an attacker(Eve) secretly intercepts and potentially alters communication between two parties. The attacker(eavesdropper) acts as a middleman in communication without the knowledge of the two parties involved. Our protocol uses $N$ decoy qubits to check for an eavesdropper, Eve. As per the protocol, Alice and Bob uses these $N$ decoy qubits in a BB84-like prepare and measure protocol. Finally, after measurement is done on both sides, they announce the bases, and matches the outcome. Further, these decoy qubits are randomly inserted in the sequence that also contains $N$ bits of information about the actual message sent by Alice. Similar to the BB84 protocol, if Eve tries to perform man-in-the-middle attack, i.e., she tries to measure the qubits sent by Alice to Bob, with the state of the decoy qubits will be changed, and Alice and Bob will find at least  $25\%$ mismatch in the outcome of the decoy qubit measurements, and she will be detected with a high probability. 

\subsection{Intercept Resend attack}
The intercept resend attack is a specific type of MITM attack in the context of quantum key distribution(QKD) where Eve intercepts the quantum state sent by Alice, measures it and resends new state(prepared according to the measurement results) to Bob. For the protocol to be secure under intercept resend attacks, only detecting presence of Eve is not sufficient, but we need to ensure that there is no leakage of secret message to Eve before she is detected. This can be done by splitting the encoded information into multiple pieces such that, for Eve to decode the message, she need access to all the pieces simultaneously. In the proposed protocol as we are checking eavesdropping after communication of every sequence of message-bits. As explained in Section~\ref{sec:4}, Alice (Bob) breaks her complete message of size $5N$ classical bits in $N$ $5$-bit pieces, and encodes them in $5$ sequences consisting $N$ qubits each. Each qubit in a sequence refers to the qubit in the same location of the five-qubit quantum state. Thus, each sequence contains only a single qubit of each $N$ pieces of the original message. This ensures Eve cannot obtain simultaneous access to more than one qubit of the entangled state. Hence the protocol is secure under intercept resend attack.

\subsection{Entangling probe attack} 
We now consider the possibility that Eve uses a black-box Unitary to entangle an ancilla state $\ket{\varepsilon}$ with qubits $2$ and $4$ in one of the quantum states sent to Bob by Alice, or vice versa. This is an entanglement based intercept-resend attack, otherwise known as the entangling probe attack. We assume that the black box unitary used by Eve is $U_E$, and it performs the transformation as provided in Eq.~\ref{eq:6}, such that, after measuring her ancilla state, Eve will know partial information about the quantum state transferred.
\begin{equation}
\label{eq:6}
    \begin{aligned}
        \ket{00}_{2,4} \ket{\varepsilon}_E \rightarrow &(\ket{00}\ket{\varepsilon_{00}} + \ket{01}\ket{\varepsilon_{01}} 
        +\ket{10}\ket{\varepsilon_{02}} + \ket{11}\ket{\varepsilon_{03}})\\
        \ket{01}_{2,4} \ket{\varepsilon}_E \rightarrow &(\ket{00}\ket{\varepsilon_{10}} + \ket{01}\ket{\varepsilon_{11}}   +\ket{10}\ket{\varepsilon_{12}} + \ket{11}\ket{\varepsilon_{13}})\\
        \ket{10}_{2,4} \ket{\varepsilon}_E \rightarrow &(\ket{00}\ket{\varepsilon_{20}} + \ket{01}\ket{\varepsilon_{21}}+ \ket{10}\ket{\varepsilon_{22}} + \ket{11}\ket{\varepsilon_{23}})\\
        \ket{11}_{2,4} \ket{\varepsilon}_E \rightarrow &(\ket{00}\ket{\varepsilon_{30}} + \ket{01}\ket{\varepsilon_{31}} + \ket{10}\ket{\varepsilon_{32}} + \ket{11}\ket{\varepsilon_{33}})
    \end{aligned}
\end{equation}
 We will show, that the stabilizer nature of LU-equivalent cluster state(s) protects the dialogue scheme against this attack, and Eve fails to get any information about the encoded message without being caught. $\ket{\psi_5}$ is a stabilizer state and its stabilizers are  S= $\lbrace X_1X_2X_3, X_3X_4X_5, Z_1Z_2, Z_4Z_5, Z_2Z_3Z_4 \rbrace$. We take these stabilizer measurement settings in a case by case manner, and discuss the security of the protocol further. 

 After Eve's interaction, the generalized state provided in Eq.~\ref{eq:1} would be, 
 \begin{equation}\label{eq:7}
 \begin{aligned}
     \ket{\psi_5}_E=& \frac{1}{2}((\ket{000}_{1,3,5}  [\ket{00}\ket{\varepsilon_{00}} + \ket{01}\ket{\varepsilon_{01}} 
 +     \ket{10}\ket{\varepsilon_{02}}+ \ket{11}\ket{\varepsilon_{03}}]_{2,4,E})\\
      & + ( \ket{001}_{1,3,5}  [\ket{00}\ket{\varepsilon_{30}} + \ket{01}\ket{\varepsilon_{31}}  + \ket{10}\ket{\varepsilon_{32}} + \ket{11}\ket{\varepsilon_{33}}]_{2,4,E})\\
      & + (\ket{110}_{1,3,5}    [\ket{00}\ket{\varepsilon_{10}} + \ket{01}\ket{\varepsilon_{11}}  + \ket{10}\ket{\varepsilon_{12}} + \ket{11}\ket{\varepsilon_{13}}]_{2,4,E})\\
      & + (\ket{101}_{1,3,5}   [\ket{00}\ket{\varepsilon_{20}} + \ket{01}\ket{\varepsilon_{21}}  + \ket{10}\ket{\varepsilon_{22}} + \ket{11}\ket{\varepsilon_{23}}]_{2,4,E})).
 \end{aligned}
 \end{equation}
Now Bob performs the stabilizer syndrome measurement on the state $\ket{\psi_5}_E$. 
Ideally, $S_i \ket{\psi_5} = \ket{\psi_5}$ for $S_i \in S$. 
\noindent \textit{Case 1:} Action of stabilizer $Z_1Z_2$
 \begin{equation}
 \label{eq:8}
     \begin{aligned}
          Z_1Z_2 \ket{\psi_5}_E &= \frac{1}{2}  ((\ket{000}_{1,3,5}[\ket{00}\ket{\epsilon_{00}}+ \ket{01}\ket{\epsilon_{01}}- \ket{10} \ket{\epsilon_{02}} - \ket{11} \ket{\epsilon_{03}}]_{2,4,E})  \\
         &+(\ket{011}_{1,3,5}[\ket{00}\ket{\epsilon_{10}}+ \ket{01}\ket{\epsilon_{11}}- \ket{10} \ket{\epsilon_{12}} - \ket{11} \ket{\epsilon_{13}}]_{2,4,E}) \\ 
         &+(\ket{110}_{1,3,5}[-\ket{00}\ket{\epsilon_{20}}- \ket{01}\ket{\epsilon_{21}}+ \ket{10} \ket{\epsilon_{22}} + \ket{11} \ket{\epsilon_{23}}]_{2,4,E}) \\
         & + (\ket{101}_{1,3,5}[-\ket{00}\ket{\epsilon_{30}}- \ket{01}\ket{\epsilon_{31}}+ \ket{10} \ket{\epsilon_{32}} + \ket{11} \ket{\epsilon_{33}}]_{2,4,E}))
     \end{aligned}
 \end{equation}
So for Eve to remain unnoticed,
\begin{equation}
    \label{eq:9}
    \begin{aligned}
        & \ket{\epsilon_{02}}= \ket{0}_{null}, \ket{\epsilon_{03}}= \ket{0}_{null}, \ket{\epsilon_{12}}= \ket{0}_{null}, \ket{\epsilon_{13}}= \ket{0}_{null}, \\
        & \ket{\epsilon_{20}}= \ket{0}_{null}, \ket{\epsilon_{21}}= \ket{0}_{null}, \ket{\epsilon_{30}}= \ket{0}_{null}, \ket{\epsilon_{31}}= \ket{0}_{null}
    \end{aligned}
\end{equation}
where $\ket{0}_{null}$ is a null ket (a vector in Hilbert space whose all components are zero).
\\ \vspace{2mm}
\noindent\textit{Case 2:} Action of stabilizer $Z_4Z_5$
 \begin{equation}
 \label{eq:10}
     \begin{aligned}
          Z_4Z_5 \ket{\psi_5}_E &= \frac{1}{2}  ((\ket{000}_{1,3,5}[\ket{00}\ket{\epsilon_{00}}- \ket{01}\ket{\epsilon_{01}}+ \ket{10} \ket{\epsilon_{02}} - \ket{11} \ket{\epsilon_{03}}]_{2,4,E})  \\
         &+(\ket{011}_{1,3,5}[-\ket{00}\ket{\epsilon_{10}}+ \ket{01}\ket{\epsilon_{11}}- \ket{10} \ket{\epsilon_{12}} +\ket{11} \ket{\epsilon_{13}}]_{2,4,E}) \\ 
         &+(\ket{110}_{1,3,5}[\ket{00}\ket{\epsilon_{20}}- \ket{01}\ket{\epsilon_{21}}+ \ket{10} \ket{\epsilon_{22}} - \ket{11} \ket{\epsilon_{23}}]_{2,4,E}) \\
         & + (\ket{101}_{1,3,5}[-\ket{00}\ket{\epsilon_{30}}+ \ket{01}\ket{\epsilon_{31}}- \ket{10} \ket{\epsilon_{32}} + \ket{11} \ket{\epsilon_{33}}]_{2,4,E}))
     \end{aligned}
 \end{equation}
So for Eve to remain unnoticed,
\begin{equation}
    \label{eq:11}
    \begin{aligned}
        & \ket{\epsilon_{01}}= \ket{0}_{null}, \ket{\epsilon_{03}}= \ket{0}_{null}, \ket{\epsilon_{10}}= \ket{0}_{null}, \ket{\epsilon_{12}}= \ket{0}_{null}, \\
        & \ket{\epsilon_{21}}= \ket{0}_{null}, \ket{\epsilon_{23}}= \ket{0}_{null}, \ket{\epsilon_{30}}= \ket{0}_{null}, \ket{\epsilon_{32}}= \ket{0}_{null}
    \end{aligned}
\end{equation}
\vspace{2 mm}
\noindent \textit{Case 3:} Action of stabilizer $Z_2Z_3Z_4$
 \begin{equation}
 \label{eq:12}
     \begin{aligned}
          Z_2Z_3Z_4 \ket{\psi_5}_E &= \frac{1}{2}  ((\ket{000}_{1,3,5}[\ket{00}\ket{\epsilon_{00}}- \ket{01}\ket{\epsilon_{01}}- \ket{10} \ket{\epsilon_{02}} + \ket{11} \ket{\epsilon_{03}}]_{2,4,E})  \\
         &+(\ket{011}_{1,3,5}[-\ket{00}\ket{\epsilon_{10}}+ \ket{01}\ket{\epsilon_{11}}+ \ket{10} \ket{\epsilon_{12}} - \ket{11} \ket{\epsilon_{13}}]_{2,4,E}) \\ 
         &+(\ket{110}_{1,3,5}[-\ket{00}\ket{\epsilon_{20}}+ \ket{01}\ket{\epsilon_{21}}+ \ket{10} \ket{\epsilon_{22}} - \ket{11} \ket{\epsilon_{23}}]_{2,4,E}) \\
         & + (\ket{101}_{1,3,5}[\ket{00}\ket{\epsilon_{30}}- \ket{01}\ket{\epsilon_{31}}- \ket{10} \ket{\epsilon_{32}} + \ket{11} \ket{\epsilon_{33}}]_{2,4,E}))
     \end{aligned}
 \end{equation}
So for Eve to remain unnoticed,
\begin{equation}
    \label{eq:13}
    \begin{aligned}
        & \ket{\epsilon_{01}}= \ket{0}_{null}, \ket{\epsilon_{02}}= \ket{0}_{null}, \ket{\epsilon_{10}}= \ket{0}_{null}, \ket{\epsilon_{13}}= \ket{0}_{null}, \\
        & \ket{\epsilon_{20}}= \ket{0}_{null}, \ket{\epsilon_{23}}= \ket{0}_{null}, \ket{\epsilon_{31}}= \ket{0}_{null}, \ket{\epsilon_{32}}= \ket{0}_{null}
    \end{aligned}
\end{equation}
\noindent \textit{Case 4:} Action of stabilizer $X_1X_2X_3$
 \begin{equation}
 \label{eq:14}
     \begin{aligned}
          X_1X_2X_3 \ket{\psi_5}_E &= \frac{1}{2}  ((\ket{110}_{1,3,5}[\ket{10}\ket{\epsilon_{00}}+ \ket{11}\ket{\epsilon_{01}}+ \ket{00} \ket{\epsilon_{02}} + \ket{01} \ket{\epsilon_{03}}]_{2,4,E})  \\
         &+(\ket{101}_{1,3,5}[\ket{10}\ket{\epsilon_{10}}+ \ket{11}\ket{\epsilon_{11}}+ \ket{00} \ket{\epsilon_{12}} + \ket{01} \ket{\epsilon_{13}}]_{2,4,E}) \\ 
         &+(\ket{000}_{1,3,5}[\ket{10}\ket{\epsilon_{20}}+ \ket{11}\ket{\epsilon_{21}}+ \ket{00} \ket{\epsilon_{22}} + \ket{01} \ket{\epsilon_{23}}]_{2,4,E}) \\
         & + (\ket{011}_{1,3,5}[\ket{10}\ket{\epsilon_{30}}+ \ket{11}\ket{\epsilon_{31}}+ \ket{00} \ket{\epsilon_{32}} + \ket{01} \ket{\epsilon_{33}}]_{2,4,E}))
     \end{aligned}
 \end{equation}
So for Eve to remain unnoticed,
\begin{equation}
    \label{eq:15}
    \begin{aligned}
        & \ket{\epsilon_{00}}= \ket{\epsilon_{22}}, \ket{\epsilon_{01}}= \ket{\epsilon_{23}}, \ket{\epsilon_{20}}= \ket{\epsilon_{02}}, \ket{\epsilon_{03}}= \ket{\epsilon_{21}}, \\
        & \ket{\epsilon_{32}}= \ket{\epsilon_{10}}, \ket{\epsilon_{11}}= \ket{\epsilon_{33}}, \ket{\epsilon_{30}}= \ket{\epsilon_{12}}, \ket{\epsilon_{13}}= \ket{\epsilon_{31}},
    \end{aligned}
\end{equation}
\noindent \textit{Case 5:} Action of stabilizer $X_3X_4X_5$
 \begin{equation}
 \label{eq:16}
     \begin{aligned}
          X_3X_4X_5 \ket{\psi_5}_E &= \frac{1}{2}  ((\ket{011}_{1,3,5}[\ket{01}\ket{\epsilon_{00}}+ \ket{00}\ket{\epsilon_{01}}+ \ket{11} \ket{\epsilon_{02}} + \ket{10} \ket{\epsilon_{03}}]_{2,4,E})  \\
         &+(\ket{000}_{1,3,5}[\ket{01}\ket{\epsilon_{10}}+ \ket{00}\ket{\epsilon_{11}}+ \ket{11} \ket{\epsilon_{12}} + \ket{10} \ket{\epsilon_{13}}]_{2,4,E}) \\ 
         &+(\ket{101}_{1,3,5}[\ket{01}\ket{\epsilon_{20}}+ \ket{00}\ket{\epsilon_{21}}+ \ket{11} \ket{\epsilon_{22}} + \ket{10} \ket{\epsilon_{23}}]_{2,4,E}) \\
         & + (\ket{110}_{1,3,5}[\ket{01}\ket{\epsilon_{30}}+ \ket{00}\ket{\epsilon_{31}}+ \ket{11} \ket{\epsilon_{32}} + \ket{10} \ket{\epsilon_{33}}]_{2,4,E}))
     \end{aligned}
 \end{equation}
So for Eve to remain unnoticed,
\begin{equation}
    \label{eq:17}
    \begin{aligned}
        & \ket{\epsilon_{00}}= \ket{\epsilon_{11}}, \ket{\epsilon_{01}}= \ket{\epsilon_{10}}, \ket{\epsilon_{13}}= \ket{\epsilon_{02}}, \ket{\epsilon_{03}}= \ket{\epsilon_{12}}, \\
        & \ket{\epsilon_{32}}= \ket{\epsilon_{23}}, \ket{\epsilon_{22}}= \ket{\epsilon_{33}}, \ket{\epsilon_{30}}= \ket{\epsilon_{21}}, \ket{\epsilon_{20}}= \ket{\epsilon_{31}},
    \end{aligned}
\end{equation}
By comparing equations \ref{eq:9},\ref{eq:11}, \ref{eq:13}, \ref{eq:15}, \ref{eq:17} we get,
\begin{equation}
    \label{eq:18}
    \begin{aligned}
        & \ket{\epsilon_{00}}= \ket{\epsilon_{11}} = \ket{\epsilon_{22}} = \ket{\epsilon_{33}} 
    \end{aligned}
\end{equation}
and $\ket{\epsilon_{ij}}= \ket{0}_{null}$ for all $i \neq j$.
Now substituting these values in \ref{eq:6} we recover the transformation performed by the black-box Unitary $U_E$ as, 
\begin{equation}
    \label{eq:19}
    \begin{aligned}
    U_E: & \ket{00}_{2,4}\ket{\varepsilon}_E \rightarrow \ket{00}_{2,4}\ket{\varepsilon_{00}}\\
    & \ket{01}_{2,4}\ket{\varepsilon}_E \rightarrow \ket{01}_{2,4}\ket{\varepsilon_{00}}\\
    & \ket{10}_{2,4}\ket{\varepsilon}_E \rightarrow \ket{10}_{2,4}\ket{\varepsilon_{00}}\\
    & \ket{11}_{2,4}\ket{\varepsilon}_E \rightarrow \ket{11}_{2,4}\ket{\varepsilon_{00}}
    \end{aligned}
\end{equation}

This proves that, as Eve has no prior information of the measurement bases chosen by Alice and Bob, the only possible Unitary she can use while staying undiscovered is not an entangling operation. Thus the joined state of Bob and Eve is a product state, and Eve cannot get any information about Alice's encoding by measuring her ancilla state. Hence the protocol is secure if there is no noise and loss in the quantum channel. 
\section{Efficiency of proposed scheme}\label{sec:6}
 In \cite{PhysRevLett.85.5635}, authors have proposed a parameter to analyze the efficiency of quantum communication protocols. 
 The parameter is defined as  $$ \eta = \frac{c}{q+b},$$
where c is the total number of classical bits exchanged (message bits), q is the total number of qubits used in protocols and b is the number of \textit{additional} classical bits required for decoding the message. This definition omits the classical bits used for checking eavesdropping.

We consider one iteration of quantum dialogue in a secure channel, i.e., Alice sends a $5-$bit message to Bob and Bob reverts another $5-$bit message back to Alice, i.e., $c=10$. Alice and Bob reuses their five-qubit (LU equivalent) cluster state, however, they uses five qubits for ancilla measurements each, i.e., q=$15$. Finally, there is no additional classical bits needed to decipher the classical message, i.e., $b=0$. This leads to $ \eta= \frac{10}{15}= 66.67\%$.

If we consider an insecure quantum channel, we need to take into account the decoy qubits need to secure the protocol against Man-in-the-middle attack, then an additional $10$ qubits are required, i.e., in this case, $\eta_ = \frac{10}{25} = 40 \%.$

\begin{table}[!hbt]
    \centering
    \resizebox{15cm}{!}{
    \begin{tabular}{| p{1cm} | p{2.2cm} | p{1.5cm} | p{5cm} | p{3cm} | p{1.4cm} | p{1.6cm} |}
    \hline
    Ref. & Assumptions & Quantum Resource & Message Flow & Security  & Efficiency $(\eta)$& Error Correction\\
    \hline
    \cite{nguyen2004quantum} & Noiseless, lossless channel & Bell states &  1. Alice encodes her message on a Bell state and sends the travel qubit. 2. Bob encodes his message on the received qubit and sends it back. 3. Alice performs measurements, decodes Bob's message and announces states and measurement outcome to Bob for decoding purpose. & It is vulnerable to intercept-resend attack and suffers information leakage. & 67\% & Not addressed\\
    \hline
    \cite{LinCY} AQD & Noiseless, lossless channel, parties pre-share a secret key K & Bell states & 1. Alice prepares and sends travel qubits from Bell states to Bob.  2. Both parties perform mutual authentication using pre-shared key and checking qubits. 3.Alice encodes her message and sends it to Bob. 4. Bob encodes his message, then performs measurement and announces it for decoding purpose. & Provides mutual authentication. Secure against impersonation, MITM, and modification attacks. Free from information leakage. & 20 \% & Not addressed \\
    \hline
    \cite{semi} SQD & Alice can perform quantum operations and Bob is restricted to perform classical operations & Bell states & 1. Alice sends travel qubits from Bell states to Bob. 2. Bob reflects some qubits for security and measures/prepares others to encode his key (KB) via XOR, then sends a permuted sequence back. 3. After checks, Alice announces her key (KA). 4. Both parties calculate the final key Kf = KA $\oplus$ KB. & Security relies on Bob's permutation of qubits, which prevents a CNOT attack from Eve. Requires hash functions to prevent manipulation attacks by Alice or Bob. & 20\% & Not addressed \\
    \hline
    \cite{yang2019new} Collective noise & Operated in collective-noise channels & Logical qubits &  1. Bob sends Alice N batches of photons, each encoded with the same message. 2. Alice selects N-1 batches for security checks. 3. Alice encodes her message on the last batch. 4. Bob reveals initial states, Alice measures and decodes. She then announces results for Bob to decode. & Secure against Trojan horse, intercept-resend, and entangle-measure attack. Does not suffer information leakage. & 22.22 \% & Robust against specific collective noises \\
    \hline 
    This work & Noiseless, lossless channel, bijection between encoding operation and ancilla outcome & LU equivalent of 5-qubit cluster state & 1. Alice encodes her message on the 5-qubit state and sends it to Bob. 2. Bob decodes Alice’s message using Non-Destructive Discrimination (NDD). 3. Bob encodes his message on the same state and sends it back. 4.  Alice decodes Bob's message using NDD. & Secure against information leakage, MITM, intercept-resend, and entangling probe attacks. & 40 \% & Single-qubit error correction scheme. \\
    \hline
    \end{tabular}}
    \caption{Comparison table with BaAn protocol, AQD protocol , SQD protocol and Collective noise protocol.}
    \label{table:7}
\end{table}

One important aspect of this dialogue scheme is the re-usability of the quantum resources. As we use non-destructive discrimination, the quantum resources needed for the protocol remains same for multiple iterations of the dialogue. i.e., if Alice and Bob wants to continue the dialogue for $2$ iterations, they can reuse all quantum registers of five qubits, increasing the efficiency of the protocol to $\frac{20}{25}$, and $\frac{20}{45}$ for the secure and insecure quantum channels respectively.  

 We provide a comparison of the efficiency of our protocol with respect to the existing literature in {Table~\ref{table:7}}. One of the distinctive property of our scheme lies in the fact that in most of the quantum dialogue protocols, one of the communicator announces initial and final states for the other communicator to decode the message, which might lead to information leakage, leading to eavesdropping. However, in our proposed protocol, by using ancilla qubits and NDD, both the communicating parties can encode and decode the message without announcing initial and final states, strengthening the security of the protocol. 

\section{Generalized protocol for n-bit classical message}
\label{sec:7}
In this work, we have proposed a scheme for two-way communication of two messages, each consisting five classical bits by utilizing the entanglement and mutual orthogonality of a set of $32$ five-qubit states that are local unitary equivalent to the five-qubit cluster state. As explained in Section~\ref{sec:4}, one can increase the number of classical bits sent through the protocol by breaking them into $N$ pieces of $5-$bit messages and parallelizing the dialogue scheme. This will increase the number of qubit used in the protocol linearly with $N$. Subsequently, one can sequentially send these $N$ pieces of $5$ classical bits using the same quantum resources, increasing the temporal complexity. 

In this section, we explore the possibility of utilizing an $n-$qubit cluster state to send an $n$-bit message. Following the protocol, the $n$-bit classical message is to be encoded on $m < n$ qubits. We now provide a generalized method to decide the value of $m$, as well as an encoding scheme. First, we divide our problem in two cases, when n is even and n is odd.

Considering $n$ is even, we assume, $n=2l$. Then, a suitable choice for $m$ is $\frac{n}{2}$. For a classical message of $n-$bits, there are $2^n$ possible combinations that can be sent in a one-way communication. As we are using Unitary encoding,  this requires $2^n= 4^l$ unitary operations to be acted upon $m$-qubits. Thus, following Section \ref{sec:3} we can design a group of Unitary operations as $\mathcal{U}_{n}$, with elements $U^{n}_e$, such that, 
$$U^{n}_e = U_G^1 \otimes U_G^2 \otimes \dots \otimes U_G^l, $$
where superscript denotes sequential ordering, and $G= \{I, X, Y, Z\}$.

When $n$ is odd then, $ m= \lceil \frac{n}{2} \rceil$, and $n+1$ is even. Thus, similar to the $n=$odd case, we can obtain the elements of the group of unitary operators of order $2^{n+1}$. Then we can replace of the groups $G$ of order $4$ with one of its subgroup of order $2$ to get the required set of operators. This prescription has also been followed in Section~\ref{sec:3} to obtain the Unitary group $\mathcal{U}_{\{A, B\}}$ for $n=5$.

Once the dense coding scheme is set, the next step is NDD of $n$-qubit cluster state. The key idea for NDD of an (LU equivalent) $n$-qubit cluster state is to break the state into smaller subspaces consisting Bell states through an unitary transformation. After the discrimination of Bell states, the inverse Unitary transformation is applied on the quantum state, restoring the original state. 

For communicating a $k$-bit classical message through the proposed quantum dialogue, the corresponding cluster state should at least have $k=n$ qubits. For any $k<n$ the scheme will work but for $k>n$, the protocol fails.\\

\begin{algorithm}
\caption{ LU equivalent of $n$-qubit cluster state.}
\textbf{Procedure:} Constructing LU equivalent of $n$-qubit cluster state.
\begin{algorithmic}
\Require $q_i$ for i = 1 to n  \Comment{$n$-qubits}
\Require $G= \lbrace I, X,Z, XZ \rbrace$
\For{$i$ in range({$n$})}
\State $h(q_i)$  \Comment{Initialize  in $\ket{+} $ state}
\State $cz(q_i,q_{i+1})$ \Comment{Controlled-Z gate}
\EndFor
\end{algorithmic}
$n$-qubit cluster state $\ket{C_n}$is formed.
\begin{algorithmic}
\For{$i$ in range($n$)}
\If{$i$ is odd}
\State $h(q_i)$ \Comment{LU Operation}
\EndIf
\EndFor
\end{algorithmic}
Initial state $\ket{\psi_n}$ is generated. \\
$\Omega = \lbrace \ket{\psi_n} \rbrace$ \Comment{Set of $2^n$ orthogonal states}
\begin{algorithmic}
\For {$i$ in range({$n$})}
\State Apply G on $q_i$  \Comment{LU Operations}
\State Add state to $\Omega$.
\EndFor
\end{algorithmic}
\textbf{End Procedure}
\end{algorithm}

\begin{algorithm}
\caption{Encoding Unitary Operation}
\textbf{Procedure}: Choosing number of encoding qubits ($m$)
\begin{algorithmic}
\Require $n$ \Comment{Number of qubits}
\If {$n$ is even}
\State $m = \frac{n}{2}$
\ElsIf {$n$ is odd}
\State $m = \lceil \frac{n}{2} \rceil$
\EndIf
\end{algorithmic}
\textbf{End Procedure}
\vspace{0.2em}

\textbf{Procedure:} Encoding unitary operations $\mathcal{U_{\{A,B\}}}$
\begin{algorithmic}
\Require $\ket{\psi_n}$   \Comment{Initial state}
\Require $\Omega$    \Comment{Set of $2^n$ orthogonal states}
\Require $S_{\ket{\psi_n}}$  \Comment{Set of stabilizers of $\ket{\psi_n}$}
\Require $G = \lbrace I, X, Z, XZ \rbrace$
\vspace{0.3em}

\noindent$\mathcal{U_{\{A,B\}}} = \lbrace\rbrace$\\
$Q_{enc} = \lbrace q_1, q_2 , \dots q_m \rbrace$  \Comment{Set of encoding qubits}
\For {$k$ in range($2^n$)}
\State $U_e^k= U_G^{q_1} \otimes U_G^{q_2} \otimes \dots \otimes U_G^{q_m}$   \Comment{$U_G \in G$}
\vspace{0.1em}

\If {$U_e^k \ket{\psi_n} \in \Omega \setminus \lbrace \mathcal{U_{\{A,B\}}} \ket{\psi_n} \rbrace$ and $U_e^k \notin S_{\ket{\psi_n}}$}
\State Add $U_e^k$ to $\mathcal{U_{\{A,B\}}}$
\EndIf
\EndFor
\vspace{0.3em}

\If{$\mathcal{|U_{\{A,B\}}}| = | \Omega|$}
\State $\mathcal{U_{\{A,B\}}}$ is the required encoding set.
\Else \hspace{0.2em} {Choose new $Q_{enc}$ and repeat the process.}
\EndIf

\end{algorithmic}
\textbf{End Procedure}
\end{algorithm}

\begin{algorithm}
\caption{Non-destructive discrimination of $\ket{\psi_n}$}
\begin{algorithmic}
\Require $\psi_n^i$  \Comment{Encoded state}
\Require $a_i$ for i= 1 to n \Comment{$n$-ancilla qubits}
\For{$i$ in range($n-1$)}
\If {$i$ is even}
\State cx$(q_i,q_{i+1})$       \Comment{Breaks the state into Bell states}
\EndIf
\EndFor
\For{$i$ in range($\lceil \frac{n}{2} \rceil$)}
\State $h(a_i)$
\State czz$(a_i,q_i,q_{i+1})$     \Comment{Calculates parity}
\State cxx$(a_{i+1},q_i,q_{i+1})$  \Comment{Calculates phase}
\EndFor
\For{$i$ in range($n-1$)}
\If {$i$ is even}
\State cx$(q_i,q_{i+1})$       \Comment{Restores the state}
\EndIf
\EndFor
\For{$i$ in range$(n)$}
\State $h(a_i)$
\State measure$(a_i)$      \Comment{Syndrome measurement}
\EndFor
\end{algorithmic}
\textbf{End procedure}
\end{algorithm}

\newpage
\section{Single Qubit Error Correction}\label{sqec}

The proposed quantum dialogue protocol is defined assuming ideal conditions (no noise). In practice, communication  never occurs under ideal scenario, but is affected by noise. Utilizing the fact that the our quantum dialogue protocol are based on a set of quantum states that are local unitary equivalent to the five-qubit cluster state, we present a single-qubit quantum error correction scheme for our protocol.

\begin{table}[!htb]
\centering
\begin{minipage}[t]{0.48\textwidth}
    \centering
    \begin{tabular}{|c|c|}
        \hline
        Error & Syndrome \\
        \hline
        $X_1$  & 00100 \\
        $Z_1$ & 10000 \\
        $XZ_1$ & 10100 \\
        \hline
        $X_2$ & 00110 \\
        $Z_2$ & 10000 \\
        $XZ_2$ & 10110 \\
        \hline
        $X_3$ & 00010 \\
        $Z_3$ & 11000 \\
        $XZ_3$ & 11010 \\
        \hline
        $X_4$ & 00011 \\
        $Z_4$ & 01000 \\
        $XZ_4$ & 01011 \\
        \hline
        $X_5$ & 00001 \\
        $Z_5$ & 01000 \\
        $XZ_5$ & 01001 \\
        \hline
    \end{tabular}
    \caption{Syndrome table for all single-qubit errors.}
    \label{table:1}
\end{minipage}
\hfill
\begin{minipage}[t]{0.48\textwidth}
    \centering
    \begin{tabular}{|c|c|}
        \hline
        Z-Error & Decoder Action \\
        \hline
        $Z_1$  &  \\
        $Z_2$  & $Z_1$ \\
        \hline
        $Z_3$ &  $Z_3$ \\
        \hline
        $Z_4$ &  \\
        $Z_5$ & $Z_4$ \\
        \hline
    \end{tabular}
    \caption{Decoder action for phase-flip error ($Z$-Error).}
    \label{table:3}
\end{minipage}
\end{table}

Two of the most frequent noises affecting the communication are bit-flip noise and phase-flip noise. Bit-flip noise changes $\ket{0}$ to $\ket{1}$ or $\ket{1}$ to $\ket{0}$ with probability $p$. Phase-flip noise changes the phase of $\ket{1}$ to -$\ket{1}$ with probability p. We explain the effects of these noises in our proposed scheme. We have only considered the single qubit noises affecting the quantum state.

\begin{figure}[!hbt]
    \Qcircuit @C=1.8em @R=1.5em {
    &  &  \mbox{Error} &  &  &  &  &  \mbox{Syndrome Measurement} &  & &  &  & \mbox{Decoder} & &\\
    &  &  \multigate{4}{E} &  \qw & \qw & \gate{X} \qwx[1] & \qw & \gate{Z} \qwx[1] & \qw & \qw & \qw & \qw & \multigate{4}{D} & \qw &\\
    &  &  \ghost{E} &  \qw & \qw & \gate{X} \qwx[1] & \qw & \gate{Z} & \gate{Z} \qwx[1] & \qw & \qw & \qw & \ghost{D} & \qw &\\
    &\lstick{\ket{\psi_5}}  &  \ghost{E} &  \qw & \qw & \gate{X} & \gate{X} \qwx[1] & \qw & \gate{Z} \qwx[1] & \qw & \qw & \qw & \ghost{D} & \qw & \rstick{\ket{\psi_5}} \\
    &  &  \ghost{E} &  \qw & \qw & \qw & \gate{X} \qwx[1] & \qw & \gate{Z} & \gate{Z} \qwx[1] & \qw & \qw & \ghost{D} & \qw & \\
    &  &  \ghost{E} &  \qw & \qw & \qw & \gate{X} & \qw & \qw & \gate{Z} & \qw & \qw & \ghost{D} & \qw &\\
    &  &  & \lstick{\ket{0}} & \gate{H} & \ctrl{-3} & \qw & \qw & \qw & \qw & \gate{H} & \meter & \cctrl{-1} \\
    &  &  & \lstick{\ket{0}} & \gate{H} & \qw & \ctrl{-2} & \qw & \qw & \qw & \gate{H} & \meter & \cctrl{-2}\\
    &  &  & \lstick{\ket{0}} & \gate{H} & \qw & \qw & \ctrl{-6} & \qw & \qw & \gate{H} & \meter & \cctrl{-3}\\
    &  &  & \lstick{\ket{0}} & \gate{H} & \qw & \qw & \qw & \ctrl{-5} & \qw & \gate{H} & \meter & \cctrl{-4}\\
    &  &  & \lstick{\ket{0}} & \gate{H} & \qw & \qw & \qw & \qw & \ctrl{-5} & \gate{H} & \meter & \cctrl{-5}
    \gategroup{2}{2}{6}{2}{0.7em}{\{}
    \gategroup{2}{14}{6}{14}{0.7em}{\}}
    \gategroup{2}{4}{11}{12}{0.7em}{--}
    \gategroup{2}{6}{4}{6}{0.7em}{--}
    \gategroup{4}{7}{6}{7}{0.7em}{--}
    \gategroup{2}{8}{3}{8}{0.7em}{--}
    \gategroup{3}{9}{5}{9}{0.7em}{--}
    \gategroup{5}{10}{6}{10}{0.7em}{--}
    }
    \caption{Single-qubit error correction mechanism for $\ket{\psi_5}$ state}
    \label{figure:20}
\end{figure}

Suppose Alice encoded the message '00100', so the transformed state is $$ \ket{\psi_5^{5}} = \frac{1}{2}(\ket{00001} + \ket{00110}+\ket{11010} + \ket{11101}$$
We assume, this state is affected by bit-flip noise on the second qubit. The affected state becomes $$\ket{\psi_5^{5}}_{E} = \frac{1}{2}(\ket{01001} + \ket{01110} + \ket{10010} + \ket{10101}).$$
When Bob performs NDD, the ancilla measurement will be '10101' which means Alice's message is '11100' which is incorrect. 
Subsequently, when Bob sends his message to Alice, wrong message is transmitted and thus our protocol becomes inefficient.

To overcome this, we provide a mechanism to detect and rectify these noises. One important aspect to consider is, that these noises can affect the communication at any point. So we need to implement error correcting mechanism at every important step to ensure noise free communication.

One of the interesting feature of cluster states and their local unitary equivalents is that they are stabilizer states. This means, there exists Unitary operations $S_i$ such that $S_i \ket{C_5} = \ket{C_5}.$ \cite{Roffe_2019} gives a brief information about stabilizers and \cite{citation-key} explains about the stabilizer code for cluster states. We use this property of cluster states to design a single-qubit error correction mechanism for our protocol.

The stabilizers of  Eq.~\ref{eq:1} are $$ S= \lbrace X_1 X_2 X_3, X_3 X_4X_5, Z_1 Z_2 , Z_4 Z_5 , Z_2 Z_3 Z_4\rbrace.$$
If we consider the single-qubit errors, we can encode the bit-flip errors as $\lbrace X_1, X_2, X_3, X_4, X_5 \rbrace$, and phase-flip errors as $\lbrace Z_1, Z_2, Z_3, Z_4, Z_5 \rbrace$, with the subscripts determining the qubit numbers. We also have the  combination of both the errors. i.e., bit-phase-flip errors as $\lbrace XZ_1, XZ_2, XZ_3, XZ_4, XZ_5 \rbrace.$ 
We observe that $\lbrace Z_1, Z_2, Z_3 \rbrace$ anti-commutes with stabilizer $X_1X_2X_3$,  $\lbrace Z_3, Z_4, Z_5 \rbrace$ anti-commutes with stabilizer $X_3X_4X_5$. 
Subsequently, $\lbrace X_1, X_2\rbrace$, $\lbrace X_4,X_5 \rbrace$, $\lbrace X_2, X_3, X_4 \rbrace$ anti-commutes with stabilizers $ Z_1 Z_2 , Z_4Z_5, Z_2 Z_3 Z_4$ respectively. Similar arguments can be given for bit-phase-flip error.
This means all the single qubit errors can be detected using the stabilizers. Table \ref{table:1} shows the syndrome measurements for all single-qubit errors.
From Table \ref{table:1}, we observe that for bit-flip error(X-error) and bit-phase-flip error(XZ-Error), we are get unique syndrome measurements. Thus, these errors can be rectified using the appropriate unitary operations.
For phase-flip error(Z-Error), we observe that the syndrome measurements are repeated. However, this does not affect our error correction scheme. We consider errors $Z_1$ and $Z_2$; both error yields the same syndrome measurement '10000'. We assume decoder action to be $D=Z_1$. If the error was $Z_1$, it is trivially nullified. In the second case, if the error was $Z_2$ then we have,
$$D \ket{\psi_5}_{error} \rightarrow Z_1 (Z_2 \ket{\psi_5}) \rightarrow Z_1 Z_2 \ket{\psi_5} \rightarrow \ket{\psi_5}, $$ 

as $Z_1 Z_2$ is one of the stabilizer of $\ket{\psi_5}$. Table \ref{table:3} shows the decoder action for phase-flip error (Z-Error).

The prescribed measurements and stabilizers works for $\ket{\psi_5}$ state. However, our protocol uses $32$ orthogonal states, and for each different state, we need to use different stabilizers. When $\ket{\psi_5}$ is acted on by an unitary $U'$,  we have, 

\begin{flalign*}
\ket{\psi_5^{'}} &= U'\ket{\psi_5} = U' S_i \ket{\psi_5}= U' S_i (U')^{\dagger} U' \ket{\psi_5}= S_i^{'} \ket{\psi_5^{'}},
\end{flalign*}

where $S_i^{'} = U^{'} S_i (U')^{\dagger}$ and $S_i \in S.$ Hence $\ket{\psi_5^{'}}$ is stabilized by $S_i^{'}$. Using above technique, we can obtain stabilizers for all the 32 orthogonal  states. 
 This error correction scheme can be applied at different stages in protocols to ensure single-qubit error free communication. Fig.~\ref{figure:20} represents an instance of the this error detection and correction scheme. 
 Therefore, our error correction scheme is able to correct all single qubit errors.
 An interesting observation is that, for single-qubit error correction in our scheme, we have not used extra resources such as repetition techniques used in quantum error correction.

 \section{Conclusion}\label{conc}
In this work, we have presented a quantum dialogue protocol using non-destructive discrimination (NDD) and a set of mutually orthogonal Local Unitary equivalent states of of five-qubit cluster state. Our protocol follows a dense coding scheme of classical messages based on a group theoretic architecture and followed by a measurement based approach for the communication protocol. We have explicitly mentioned the dense coding scheme, and shown that protocol ensures security against common quantum attacks. We computed the efficiency and computational complexity of the proposal, and provided simple generalization techniques to transmit n-bit classical message. Finally, using stabilizer properties of cluster state and its LU equivalents, we have explicitly outlined a  single qubit error correction scheme for our protocol. 
The most interesting aspect of our protocol is that, after one round of two-way communication, the underlying cluster state is not destroyed and it is in one of the 32 orthogonal states $\ket{\psi_5^i}$. If Alice applies inverse unitary operations ($U_A,U_B$) on this state, we get the initial state $\ket{\psi_5}$. NDD allows the re-usability of the state for further communication, without needing any extra resource. This is the major advantage of this scheme over the existing quantum dialogue protocols. In practical scenario, the re-usability might be affected by noise, decoherence etc. We plan to examine the re-usability aspect of the protocol for practical communication purposes.



\bibliographystyle{ieeetr}
\bibliography{ref}

@article{bennett2014quantum,
  title={Quantum cryptography: Public key distribution and coin tossing},
  author={Bennett, Charles H and Brassard, Gilles},
  journal={Theoretical computer science},
  volume={560},
  pages={7--11},
  year={2014},
  publisher={Elsevier}
}

@article{ekert1991quantum,
  title={Quantum cryptography based on Bell’s theorem},
  author={Ekert, Artur K},
  journal={Physical review letters},
  volume={67},
  number={6},
  pages={661},
  year={1991},
  publisher={APS}
}

@article{bennett1992quantum,
  title={Quantum cryptography using any two nonorthogonal states},
  author={Bennett, Charles H},
  journal={Physical review letters},
  volume={68},
  number={21},
  pages={3121},
  year={1992},
  publisher={APS}
}

@article{goldenberg1995quantum,
  title={Quantum cryptography based on orthogonal states},
  author={Goldenberg, Lior and Vaidman, Lev},
  journal={Physical Review Letters},
  volume={75},
  number={7},
  pages={1239},
  year={1995},
  publisher={APS}
}

@article{zhou2004quantum,
  title={Quantum key distribution in 50-km optic fibers},
  author={Zhou, Chunyuan and Wu, Guang and Chen, Xiuliang and Li, Hexiang and Zeng, Heping},
  journal={Science in China Series G: Physics and Astronomy},
  volume={47},
  pages={182--188},
  year={2004},
  publisher={Springer}
}

@article{wang2005experimental,
  title={Experimental realization of quantum cryptography communication in free space},
  author={Wang, Chuan and Zhang, Jingfu and Wang, Pingxiao and Deng, Fuguo and Al, Qing and Long, Guilu},
  journal={Science in China Series G: Physics Mechanics and Astronomy},
  volume={48},
  pages={237--246},
  year={2005},
  publisher={Springer}
}

@article{long2002theoretically,
  title={Theoretically efficient high-capacity quantum-key-distribution scheme},
  author={Long, Gui-Lu and Liu, Xiao-Shu},
  journal={Physical Review A},
  volume={65},
  number={3},
  pages={032302},
  year={2002},
  publisher={APS}
}

@article{deng2003two,
  title={Two-step quantum direct communication protocol using the Einstein-Podolsky-Rosen pair block},
  author={Deng, Fu-Guo and Long, Gui Lu and Liu, Xiao-Shu},
  journal={Physical Review A},
  volume={68},
  number={4},
  pages={042317},
  year={2003},
  publisher={APS}
}

@article{lee2006quantum,
  title={Quantum direct communication with authentication},
  author={Lee, Hwayean and Lim, Jongin and Yang, HyungJin},
  journal={Physical Review A—Atomic, Molecular, and Optical Physics},
  volume={73},
  number={4},
  pages={042305},
  year={2006},
  publisher={APS}
}

@article{cao2010quantum,
  title={Quantum secure direct communication with cluster states},
  author={Cao, WeiFeng and Yang, YuGuang and Wen, QiaoYan},
  journal={Science China Physics, Mechanics and Astronomy},
  volume={53},
  pages={1271--1275},
  year={2010},
  publisher={Springer}
}

@article{zhu2017experimental,
  title={Experimental long-distance quantum secure direct communication},
  author={Zhu, Feng and Zhang, Wei and Sheng, Yubo and Huang, Yidong},
  journal={Science Bulletin},
  volume={62},
  number={22},
  pages={1519--1524},
  year={2017},
  publisher={Elsevier}
}

@article{nguyen2004quantum,
  title={Quantum dialogue},
  author={Nguyen, Ba An},
  journal={Physics Letters A},
  volume={328},
  number={1},
  pages={6--10},
  year={2004},
  publisher={Elsevier}
}

@article{zhong2005quantum,
  title={Quantum dialogue revisited},
  author={Zhong-Xiao, Man and Zhan-Jun, Zhang and Yong, Li},
  journal={Chinese Physics Letters},
  volume={22},
  number={1},
  pages={22},
  year={2005},
  publisher={IOP Publishing}
}

@article{an2005secure,
  title={Secure dialogue without a prior key distribution},
  author={An, Nguyen Ba},
  journal={Journal-Korean Physical Society},
  volume={47},
  number={4},
  pages={562},
  year={2005},
  publisher={Korean Physical Society; 1999}
}

@article{xia2006quantum,
  title={Quantum dialogue by using the GHZ state},
  author={Xia, Yan and Fu, Chang-Bao and Zhang, Shou and Hong, Suc-Kyoung and Yeon, Kyu-Hwang and Um, Chung-In},
  journal={arXiv preprint quant-ph/0601127},
  year={2006}
}

@article{li2009quantum,
  title={Quantum dialogue protocol using a class of three-photon W states},
  author={Li, Dong and Xiao-Ming, Xiu and Ya-Jun, Gao and Feng, Chi},
  journal={Communications in Theoretical Physics},
  volume={52},
  number={5},
  pages={853},
  year={2009},
  publisher={IOP Publishing}
}

@article{wen2007secure,
  title={Secure quantum telephone},
  author={Wen, Xiaojun and Liu, Yun and Zhou, Nanrun},
  journal={Optics communications},
  volume={275},
  number={1},
  pages={278--282},
  year={2007},
  publisher={Elsevier}
}

@article{chauhan2021quantum,
  title={QUANTUM DIALOGUE PROTOCOL USING SIX QUBIT CLUSTER STATES WITH OPTIMAL SUPERDENSE CODING.},
  author={Chauhan, S and Gupta, NL},
  journal={International Journal on Information Technologies \& Security},
  volume={13},
  number={4},
  year={2021}
}

@article{dong2008controlled,
  title={A controlled quantum dialogue protocol in the network using entanglement swapping},
  author={Dong, Li and Xiu, Xiao-Ming and Gao, Ya-Jun and Chi, Feng},
  journal={Optics Communications},
  volume={281},
  number={24},
  pages={6135--6138},
  year={2008},
  publisher={Elsevier}
}

@article{gao2010two,
  title={Two quantum dialogue protocols without information leakage},
  author={Gao, Gan},
  journal={Optics communications},
  volume={283},
  number={10},
  pages={2288--2293},
  year={2010},
  publisher={Elsevier}
}

@article{ye2013quantum,
  title={Quantum dialogue without information leakage based on the entanglement swapping between any two Bell states and the shared secret Bell state},
  author={Ye, Tian-Yu and Jiang, Li-Zhen},
  journal={Physica Scripta},
  volume={89},
  number={1},
  pages={015103},
  year={2013},
  publisher={IOP Publishing}
}

@article{wang2016efficient,
  title={Efficient quantum dialogue using entangled states and entanglement swapping without information leakage},
  author={Wang, He and Zhang, Yu Qing and Liu, Xue Feng and Hu, Yu Pu},
  journal={Quantum Information Processing},
  volume={15},
  pages={2593--2603},
  year={2016},
  publisher={Springer}
}

@article{xin2006secure,
  title={Secure quantum dialogue based on single-photon},
  author={Xin, Ji and Shou, Zhang},
  journal={Chinese Physics},
  volume={15},
  number={7},
  pages={1418},
  year={2006},
  publisher={IOP Publishing}
}

@article{shi2010quantum,
  title={Quantum secure dialogue by using single photons},
  author={Shi, Guo-Fang and Xi, Xiao-Qiang and Hu, Ming-Liang and Yue, Rui-Hong},
  journal={Optics communications},
  volume={283},
  number={9},
  pages={1984--1986},
  year={2010},
  publisher={Elsevier}
}

@article{maitra2017measurement,
  title={Measurement device-independent quantum dialogue},
  author={Maitra, Arpita},
  journal={Quantum Information Processing},
  volume={16},
  pages={1--15},
  year={2017},
  publisher={Springer}
}

@article{das2020two,
  title={Two efficient measurement device independent quantum dialogue protocols},
  author={Das, Nayana and Paul, Goutam},
  journal={International Journal of Quantum Information},
  volume={18},
  number={07},
  pages={2050038},
  year={2020},
  publisher={World Scientific}
}

@article{zhang2024measurement,
  title={Measurement-device-independent quantum dialogue based on entanglement swapping and phase encoding},
  author={Zhang, Cheng and Zhou, Lan and Zhong, Wei and Du, Ming-Ming and Sheng, Yu-Bo},
  journal={Quantum Information Processing},
  volume={23},
  number={2},
  pages={52},
  year={2024},
  publisher={Springer}
}

@article{gao2008revisiting,
  title={Revisiting the security of quantum dialogue and bidirectional quantum secure direct communication},
  author={Gao, Fei and Guo, FenZhuo and Wen, QiaoYan and Zhu, FuChen},
  journal={Science in China Series G: Physics, Mechanics and Astronomy},
  volume={51},
  number={5},
  pages={559--566},
  year={2008},
  publisher={Springer}
}

@article{tan2008classical,
  title={Classical correlation in quantum dialogue},
  author={Tan, Yong-gang and Cai, Qing-Yu},
  journal={International Journal of Quantum Information},
  volume={6},
  number={02},
  pages={325--329},
  year={2008},
  publisher={World Scientific}
}

@article{yang2013quantum,
  title={Quantum dialogue protocols immune to collective noise},
  author={Yang, Chun-Wei and Hwang, Tzonelih},
  journal={Quantum information processing},
  volume={12},
  pages={2131--2142},
  year={2013},
  publisher={Springer}
}

@article{ye2014information,
  title={Information leakage resistant quantum dialogue against collective noise},
  author={Ye, TianYu},
  journal={Science China Physics, Mechanics \& Astronomy},
  volume={57},
  pages={2266--2275},
  year={2014},
  publisher={Springer}
}

@article{ye2022fault,
  title={Fault tolerant channel-encrypting quantum dialogue against collective noise},
  author={Ye, Tian-Yu},
  journal={arXiv preprint arXiv:2205.03223},
  year={2022}
}

@article{yang2019new,
  title={New secure quantum dialogue protocols over collective noisy channels},
  author={Yang, Yu-Guang and Gao, Shang and Zhou, Yi-Hua and Shi, Wei-Min},
  journal={International Journal of Theoretical Physics},
  volume={58},
  pages={2810--2822},
  year={2019},
  publisher={Springer}
}

@article{banerjee2017asymmetric,
  title={Asymmetric quantum dialogue in noisy environment},
  author={Banerjee, Anindita and Shukla, Chitra and Thapliyal, Kishore and Pathak, Anirban and Panigrahi, Prasanta K},
  journal={Quantum Information Processing},
  volume={16},
  pages={1--23},
  year={2017},
  publisher={Springer}
}

@article{shukla2013group,
  title={On the group-theoretic structure of a class of quantum dialogue protocols},
  author={Shukla, Chitra and Kothari, Vivek and Banerjee, Anindita and Pathak, Anirban},
  journal={Physics Letters A},
  volume={377},
  number={7},
  pages={518--527},
  year={2013},
  publisher={Elsevier}
}

@article{jain2009secure,
  title={Secure quantum conversation through non-destructive discrimination of highly entangled multipartite states},
  author={Jain, Sakshi and Muralidharan, Sreraman and Panigrahi, Prasanta K},
  journal={Europhysics letters},
  volume={87},
  number={6},
  pages={60008},
  year={2009},
  publisher={IOP Publishing}
}

@article{Samal_2010,
doi = {10.1088/0953-4075/43/9/095508},
url = {https://dx.doi.org/10.1088/0953-4075/43/9/095508},
year = {2010},
month = {apr},
publisher = {},
volume = {43},
number = {9},
pages = {095508},
author = {Samal, Jharana Rani and Gupta, Manu and Panigrahi, P K and Kumar, Anil},
title = {Non-destructive discrimination of Bell states by NMR using a single ancilla qubit},
journal = {Journal of Physics B: Atomic, Molecular and Optical Physics},
}

@article{article,
author = {Satyajit, Saipriya and Srinivasan, Karthik and Behera, Bikash and Panigrahi, Prasanta},
year = {2018},
month = {07},
pages = {},
title = {Nondestructive discrimination of a new family of highly entangled states in IBM quantum computer},
volume = {17},
journal = {Quantum Information Processing},
doi = {10.1007/s11128-018-1976-9}
}

@article{SISODIA20173860,
title = {Experimental realization of nondestructive discrimination of Bell states using a five-qubit quantum computer},
journal = {Physics Letters A},
volume = {381},
number = {46},
pages = {3860-3874},
year = {2017},
issn = {0375-9601},
doi = {https://doi.org/10.1016/j.physleta.2017.09.050},
url = {https://www.sciencedirect.com/science/article/pii/S0375960117309702},
author = {Mitali Sisodia and Abhishek Shukla and Anirban Pathak},
}

@article{articleLu,
author = {Lu, Chao-Yang and Zhou, Xiaoqi and Gühne, Otfried and Gao, Weibo and Zhang, Jin and Yuan, Zhen-Sheng and Goebel, Alexander and Yang, Tao and Pan, Jian-Wei},
year = {2006},
month = {10},
pages = {},
title = {Experimental entanglement of six photons in graph states},
volume = {3},
journal = {Nature Physics},
doi = {10.1038/nphys507}
}

@article{Blythe_2006,
doi = {10.1088/1367-2630/8/10/231},
url = {https://dx.doi.org/10.1088/1367-2630/8/10/231},
year = {2006},
month = {oct},
publisher = {},
volume = {8},
number = {10},
pages = {231},
author = {Blythe, P J and Varcoe, B T H},
title = {A cavity-QED scheme for cluster-state quantum computing using crossed atomic beams},
journal = {New Journal of Physics},
}

@article{PhysRevA.78.062333,
  title = {Quantum-information splitting using multipartite cluster states},
  author = {Muralidharan, Sreraman and Panigrahi, Prasanta K.},
  journal = {Phys. Rev. A},
  volume = {78},
  issue = {6},
  pages = {062333},
  numpages = {5},
  year = {2008},
  month = {Dec},
  publisher = {American Physical Society},
  doi = {10.1103/PhysRevA.78.062333},
  url = {https://link.aps.org/doi/10.1103/PhysRevA.78.062333}
}

@article{articleHaddadi,
author = {Abhijeet, K. and Haddadi, Saeed and Pourkarimi, Mohammad and Behera, Bikash and Panigrahi, Prasanta},
year = {2020},
month = {08},
pages = {13608},
title = {Experimental realization of controlled quantum teleportation of arbitrary qubit states via cluster states},
volume = {10},
journal = {Scientific Reports},
doi = {10.1038/s41598-020-70446-8}
}

@article{Roffe_2019,
   title={Quantum error correction: an introductory guide},
   volume={60},
   ISSN={1366-5812},
   url={http://dx.doi.org/10.1080/00107514.2019.1667078},
   DOI={10.1080/00107514.2019.1667078},
   number={3},
   journal={Contemporary Physics},
   publisher={Informa UK Limited},
   author={Roffe, Joschka},
   year={2019},
   month=jul, pages={226–245} }

@article{citation-key,
    author = {Song, Dan and Cao, Zhengwen and Zhang, Shuanghao and Feng, Jie and Li, Yan and Chai, Geng},
    title = {A quantum stabilizer code associated with cluster states},
    journal ={} ,
    year = {2017}}

@article{PhysRevLett.85.5635,
  title = {Quantum Key Distribution in the Holevo Limit},
  author = {Cabello, Ad\'an},
  journal = {Phys. Rev. Lett.},
  volume = {85},
  issue = {26},
  pages = {5635--5638},
  numpages = {0},
  year = {2000},
  month = {Dec},
  publisher = {American Physical Society},
  doi = {10.1103/PhysRevLett.85.5635},
  url = {https://link.aps.org/doi/10.1103/PhysRevLett.85.5635}
}

@article{Briege,
  title = {A One-Way Quantum Computer},
  author = {Raussendorf, Robert and Briegel, Hans J.},
  journal = {Phys. Rev. Lett.},
  volume = {86},
  issue = {22},
  pages = {5188--5191},
  numpages = {0},
  year = {2001},
  month = {May},
  publisher = {American Physical Society},
  doi = {10.1103/PhysRevLett.86.5188},
  url = {https://link.aps.org/doi/10.1103/PhysRevLett.86.5188}
}

@INPROCEEDINGS{9605278,
  author={Mandal, Arijit and Banerjee, Shreya and Panigrahi, Prasanta K.},
  booktitle={2021 IEEE International Conference on Quantum Computing and Engineering (QCE)}, 
  title={Quantum Image Representation on Clusters}, 
  year={2021},
  volume={},
  number={},
  pages={89-99},
  doi={10.1109/QCE52317.2021.00025}}

@article{PhysRevA.111.012603,
  title = {Device-independent quantum secret sharing with advanced random key generation basis},
  author = {Zhang, Qi and Ying, Jia-Wei and Wang, Zhong-Jian and Zhong, Wei and Du, Ming-Ming and Shen, Shu-Ting and Li, Xi-Yun and Zhang, An-Lei and Gu, Shi-Pu and Wang, Xing-Fu and Zhou, Lan and Sheng, Yu-Bo},
  journal = {Phys. Rev. A},
  volume = {111},
  issue = {1},
  pages = {012603},
  numpages = {14},
  year = {2025},
  month = {Jan},
  publisher = {American Physical Society},
  doi = {10.1103/PhysRevA.111.012603},
  url = {https://link.aps.org/doi/10.1103/PhysRevA.111.012603}
}

@misc{javadiabhari2024quantumcomputingqiskit,
      title={Quantum computing with Qiskit}, 
      author={Ali Javadi-Abhari and Matthew Treinish and Kevin Krsulich and Christopher J. Wood and Jake Lishman and Julien Gacon and Simon Martiel and Paul D. Nation and Lev S. Bishop and Andrew W. Cross and Blake R. Johnson and Jay M. Gambetta},
      year={2024},
      eprint={2405.08810},
      archivePrefix={arXiv},
      primaryClass={quant-ph},
      url={https://arxiv.org/abs/2405.08810}, 
}

@article{CIQD,
    author = {Long Zhang and Shu Dong and Ke-Jia Zhang and Hong-Wei Sun },
    title = {A Controller-Independent Quantum Dialogue Protocol with Four-Particle States},
    journal = {International Journal of Theoretical Physics} ,
    volume = {58},
    issue = {6},
    pages= {1927-1936},
    year = {2019},
}

@article{Shukla2025,
    author = {Chitra Shukla and Abhishek Shukla and Symeon Chatzinotas and Milos Nesladek} ,
    title = {Orthogonal-state-based measurement device independent quantum communication: a noise-resilient approach} ,
    journal ={AAPPS Bulletin} ,
    year = {2025},
    volume= {35},
    number= {20},
}

@article{shend,
    author = {Shen, D. and Ma, W. and Yin, X. et al.} ,
    title = {Quantum Dialogue with Authentication Based on Bell States},
    journal = {Int J Theor Phys },
    year ={2013},
    pages = {1825-1835},
}

@article{LinCY,
    author = {Lin, CY. and Yang, CW. and Hwang, T. },
    title = {Authenticated Quantum Dialogue Based on Bell States},
    journal = {Int J Theor Phys },
    year = {2015},
}

@article{prob,
    author = {Hwang, T.and  Luo, YP.} ,
    title = {Probabilistic authenticated quantum dialogue },
    journal = {Quantum Inf Process},
    pages={4631–4650 },
    year = {2015},
}

@article{Tian,
    author = {Ye, Tian-Yu and Ye, Chong-Qiang},
    title = {Semi-quantum Dialogue Based on Single Photons},
    journal = {International Journal of Theoretical Physics},
    volume={57},
    year = {2018},
}

@article{semi,
    author = {Shukla, C. and Thapliyal, K. and  Pathak, A.},
    title = {Semi-quantum communication: protocols for key agreement, controlled secure direct communication and dialogue},
    journal = {Quantum Inf Process},
    year = {2017},
}

@article{zhen-zhen,
    author = {Zhen-Zhen Li and Run-Ze He and Zhen-Zhen Zhang and Hai-Yang Ding and Dong-Fei Wang},
    title = {Semi-quantum dialogue protocol based on four-particle $\Omega$ state},
    journal = {Chinese Journal of Physics},
    volume={95},
    pages={348-357},
    year = {2025}
}

@article{Hzu,
    author = {Chang, CH. and Yang, CW. and Hzu, GR. et al.},
    title = {Quantum dialogue protocols over collective noise using entanglement of GHZ state},
    journal = {Quantum Inf Process},
    pages={2971–2991},
    year = {2016},
}

@article{two-step,
    author = {Lin, J. and Chang, CY. and Tsai, CW. et al.},
    title = {Two-step quantum dialogue protocols against collective noises},
    journal = { EPJ Quantum Technol},
    year = {2024},
}






\appendix
\section{Appendix}\label{Appen}

\textbf{1. Group formation of Encoding Unitaries:} We discuss the group structure for the encoding unitary operations. First let us consider the set $G= \lbrace I, X, Z, XZ \rbrace$ with multiplication rule. The multiplication rule is defined as bit-wise dot product such that the global phase is ignored. $I$ is the identity element of $G$. The elements of $G$ are Pauli matrices hence they satisfy associativity. We observe that, multiplying any two elements of the set yields another element. Also as the elements are Pauli matrices, they are self-inverse. Hence, G under the multiplication rule forms a group of order 4. As the global phase is ignored, $G$ is an Abelian group.
Let $S_G = \lbrace I, X \rbrace$ and $S_G \subset G$. Under the multiplication rule, $S_G$ forms a group and thus it is a subgroup of G of order 2.\\

Now, considering $\mathcal{U_{\{A,B\}}}$ as the set of all encoding unitary operations with the multiplication rule, any element $U_e \in \mathcal{U_{\{A,B\}}}$ can be expressed as 
$$ U_e = U_G \otimes U_G \otimes U_{S_G}$$
where $U_G$ is an element of group $G$ and $U_{S_G}$ is an element of group $S_G$. Since $I$ is the identity element of groups $G$ and $S_G$, $I_1 \otimes I_3 \otimes I_5$ is identity element of $\mathcal{U_{\{A,B\}}}$. The entries of $\mathcal{U_{\{A,B\}}}$ are tensor product of Pauli matrices and thus satisfy associativity. Let $U_e, U'_e \in \mathcal{U_{\{A,B\}}}$, then we get
\begin{equation}
    \begin{aligned}
        U_e.U'_e &= (U_G \otimes U_G \otimes U_{S_G}).(U'_G \otimes U'_G \otimes U'_{S_G}) \\&
        = (U_G .U'_G) \otimes
        (U_G . U'_G) \otimes (U_{S_G} . U'_{S_G}) \\& = U''_G \otimes U''_G \otimes U''_{S_G} \in \mathcal{U_{\{A,B\}}} 
    \end{aligned}
\end{equation}
where $U''_{G}= U_G .U'_G $ is an element of $G$ and $U''_{S_G} = U_{S_G} . U'_{S_G} $ is an element of $S_G$. Hence, we observe that multiplying any two elements of the set yields another element of the set. Now, consider
\begin{equation}
    \begin{aligned}
        U_e.U_e &= (U_G \otimes U_G \otimes U_{S_G}). (U_G \otimes U_G \otimes U_{S_G})\\ &= (U_G .U_G) \otimes (U_G \otimes U_G) \otimes (U_{S_G}.U_{S_G}) \\ & = I \otimes I \otimes I = I_1 \otimes I_3 \otimes I_5 \in \mathcal{U_{\{A,B\}}}
    \end{aligned}
\end{equation}
Using the property that the elements of $G$ and $S_G$ are self-inverse, we observe that elements of $\mathcal{U_{\{A,B\}}} $ are also self-inverse. Therefore, the set $\mathcal{U_{\{A,B\}}} $ under the multiplication rule forms a group.
\vspace{2em}

\noindent \textbf{2.Proof for uniqueness of ancilla syndrome per message:} To prove the bijection between the classical message and ancilla syndrome, we define two functions $f$ and $g$. We show that both the functions are bijective and thus $g \circ f $ is also bijective. \\
\vspace{.8em}

\textbf{Classical message $\xrightarrow[Encoding]{f}$ Orthogonal State $\xrightarrow[NDD]{g}$ Ancilla Outcome}
\vspace{.8em}

We observe that all three sets have the same cardinality (n=32). Consider the  function $g$ that maps an orthogonal state to an ancilla outcome. The action of this function is same as performing NDD on an orthogonal state. We have already observed that performing NDD on an orthogonal state gives unique ancilla outcome. As the cardinality of both the sets is same, NDD is bijection and hence, $g$ is a bijective map.\\
\noindent Now, let us consider the function $f$. The function  $f: \{0,1 \} ^{\otimes 5} \rightarrow \Omega$ is defined as $$ f(m) = U_e^m \ket{\psi_5}$$
where $m=m_1m_2m_3m_4m_5$. The action of this function is same as encoding our message. 
Let $$f(m)=f(n) \implies U_e^m \ket{\psi_5} = U_e^n \ket{\psi_5}$$ Multiplying ${U_e^n}^\dagger$ on both sides we get, 
$$ {U_e^n}^\dagger U_e^m \ket{\psi_5} = {U_e^n}^\dagger U_e^n \ket{\psi_5} = I$$
The encoding set $\mathcal{U_{\{A,B\}}} $ is chosen such that the non-identity unitary operations are not stabilizers of $\ket{\psi_5}$.
$$\mathcal{U_{\{A,B\}}} \cap S_{\ket{\psi_5}} = I  $$
Also as each element of group $\mathcal{U_{\{A,B\}}} $ is self inverse, we have,
$$ U_e^m= (U_e^n)\dagger = U_e^n$$
$$ \implies m=n$$
Hence $f$ is injective.\\
Now, for any orthogonal state $\ket{\psi_5^i} \in \Omega$, we have,
$$ \ket{\psi_5^i}= U_e^{m'} \ket{\psi_5}$$
Hence for every $y = \ket{\psi_5^i} \in \Omega$, $\exists$ $x = m' \in \{0,1 \}^{\otimes 5}$ such that $f(x)=y$.
Thus, $f$ is surjective and hence, $f$ is a bijective function.\\
Since, $f$ and $g$ both are bijective, $g \circ f$ is also bijective. This proves the uniquness between the classical message and the ancilla outcome.
\vspace{2cm}
\begin{table}[!hbt]
    \centering
    \resizebox{13cm}{!}{
    \begin{tabular}{|c|c|}
    \hline
    Notations & Summary\\ 
    \hline
    $\ket{C_n}$ & $n$-qubit cluster state \\
    \hline
    $\ket{\psi_n}$ & LU-equivalent of $n$-qubit cluster state \\
    \hline
    $S_{\ket{\psi_n}}$ & Stabilizers of $\ket{\psi_n}$\\
    \hline
    $X$ $(\sigma_x)$ & Pauli X-operator \\
    \hline
    $Z$ $(\sigma_z)$ & Pauli Z-operator \\
    \hline
    $U_e$ & Encoding unitary operation \\
    \hline
    $\mathcal{U_{\{A,B\}}} $ & Set of all encoding operation \\ 
    \hline
    $\Omega$ & Set of $2^n$ mutually orthogonal LU equivalent cluster states\\
    \hline
    $\eta$ & Efficiency parameter\\
    \hline
    $h$ & Hadamard gate \\
    \hline
    cxx & Controlled-Not-Not gate ( 1 control qubit, 2 target qubits) \\
    \hline
    czz & Controlled-Z-Z gate (1 Control qubit, 2 target qubits)\\
    \hline
    \end{tabular}}
    \caption{Notation table}
    \label{tab:22}
\end{table}

\vspace{4em}


\end{document}